\documentclass[aps,prb,a4paper,twocolumn,floatfix,showpacs,superscriptaddress,preprintnumbers,10pt]{revtex4-2}
\usepackage{amsmath}
\usepackage{amssymb}
\usepackage{graphicx}
\usepackage{xcolor}
\definecolor{gray}{rgb}{0.75, 0.75, 0.75}
\usepackage{upgreek}
\usepackage[utf8]{inputenc}
\usepackage[normalem]{ulem}
\usepackage{hyperref}
\usepackage{orcidlink}
\usepackage{siunitx}
\usepackage{layouts}
\usepackage[percent]{overpic}

\begin{document}

\title{Phonon emission due to dc-driven electron heating in a two-dimensional electron gas containing artificial scatterers}

\author{Frederik Bartels\,\orcidlink{0009-0001-9903-8265}}
\affiliation{Institut f\"ur Experimentelle Physik der kondensierten Materie, Heinrich-Heine-Universit\"at, Universit\"atsstra\ss e 1, 40225 D\"usseldorf, Germany}

\author{Mihai Cerchez\,\orcidlink{0000-0003-3807-6708}}
\affiliation{Institut f\"ur Experimentelle Physik der kondensierten Materie, Heinrich-Heine-Universit\"at, Universit\"atsstra\ss e 1, 40225 D\"usseldorf, Germany}

\author{Jan Lange\,\orcidlink{0009-0002-9475-6169}}
\affiliation{Institut f\"ur Experimentelle Physik der kondensierten Materie, Heinrich-Heine-Universit\"at, Universit\"atsstra\ss e 1, 40225 D\"usseldorf, Germany}

\author{Dominique Mailly\,\orcidlink{0000-0002-2174-8785}}
\affiliation{Universit\'e Paris-Saclay, CNRS, Centre de Nanosciences et de Nanotechnologies, 91120, Palaiseau, France}

\author{Klaus Pierz\,\orcidlink{0000-0001-6400-1799}}
\affiliation{Physikalisch-Technische Bundesanstalt, Bundesallee 100, 38116 Braunschweig, Germany}

\author{Hans W. Schumacher\,\orcidlink{0000-0001-7100-682X}}
\affiliation{Physikalisch-Technische Bundesanstalt, Bundesallee 100, 38116 Braunschweig, Germany}

\author{Thomas Heinzel\,\orcidlink{0000-0001-5149-4379}}\email{Thomas.Heinzel@hhu.de}
\affiliation{Institut f\"ur Experimentelle Physik der kondensierten Materie, Heinrich-Heine-Universit\"at, Universit\"atsstra\ss e 1, 40225 D\"usseldorf, Germany}

\date{\today}

\begin{abstract}
We report the observation of phonon-induced resistance oscillations (PIROs) up to order $\varepsilon_\mathrm{{ph}} = 8$ in a two-dimensional electron gas containing artificial scatterers, activated by electron heating due to an applied dc current which enables phonon emission. Under equilibrium conditions, only two weak PIRO maxima are detectable. The dc current simultaneously generates Hall field induced resistance oscillations (HIROs). A precise analysis of their amplitude, which is modulated by the double resonance with PIROs, dramatically enhances the visibility of the latter compared to the equilibrium case. We identify transverse acoustic phonons with a sound velocity of $s = 3.4$\,km/s as the relevant mode and analyze the oscillatory amplitude modulation in the subsonic regime ($\varepsilon_{\mathrm{dc}} < \varepsilon_{\mathrm{ph}}$) within the theoretical framework of Dmitriev \textit{et al.} At the sound barrier ($\varepsilon_{\mathrm{dc}} = \varepsilon_{\mathrm{ph}}$), where the electron drift velocity reaches the speed of sound, we observe a phase shift in the oscillatory electron-phonon interaction that independently confirms recent experimental findings.
\end{abstract}

\maketitle

\section{Introduction} \label{SecIntroduction}

Phonon-induced resistance oscillations (PIROs) are $1/B$-periodic oscillations in the differential magnetoresistivity $r_{xx}(B)$ arising from the scattering of two-dimensional electrons at acoustic phonons~\cite{Gurevich1961,Yang2002b,Bykov2005,Zhang2004b,Hatke2009,Bykov2009,Bykov2010,Hatke2011,Kumaravadivel2019,Greenaway2019}. In contrast to magnetophonon resonances involving optical phonons~\cite{Gurevich1961, Tsui1980, Englert1982,Kido1982}, PIROs are observed in a temperature range of roughly 10\,K and occur when the energy of an acoustic phonon with momentum transfer $2k_\mathrm{F}$ matches the Landau level spacing. The resonance condition is $\varepsilon_{\mathrm{ph}} = \omega_\mathrm{s}/\omega_\mathrm{c} \in \mathbb{N}$, where $\omega_\mathrm{c}$ is the cyclotron frequency and $\omega_\mathrm{s} = 2k_\mathrm{F}s$ with $k_\mathrm{F}$ the Fermi wave number and $s$ the sound velocity. PIROs have been observed in GaAs quantum wells with high electron densities and mobilities~\cite{Hatke2009,Zhang2008b,Wang2026}, as well as in graphene~\cite{Kumaravadivel2019,Greenaway2019}, but typically require elevated temperatures to thermally populate the relevant phonon modes.

A related class of $1/B$-periodic oscillations are Hall field induced resistance oscillations (HIROs)~\cite{Yang2002,Bykov2005b,Zhang2007,Zhang2007b,Vavilov2007,Lei2007,Bykov2008,Vitkalov2009,Hatke2009b,Hatke2011b,Bykov2012,Hatke2012b,Studenikin2012,Dmitriev2012,Shi2014c,Shi2017,Zudov2017,Greenaway2021,Wang2023,Bartels2025,Wang2026,Bartels2026b}, which occur in the differential longitudinal resistivity as well and have their origin in elastic electron scattering between Landau levels tilted by the Hall field~\cite{Yang2002,Vavilov2007}. Their resonance condition is $\varepsilon_{\mathrm{dc}} = \omega_\mathrm{H}/\omega_\mathrm{c} \in \mathbb{N}$, where $\hbar\omega_\mathrm{H} = eE(2r_\mathrm{c})$ is the Hall energy, with the Hall electric field $E$ and the cyclotron radius $r_\mathrm{c}$. When the temperature reaches a regime where acoustic phonon scattering becomes significant, the conditions for observing HIROs and PIROs can be met simultaneously. This results in a double resonance: the amplitude of the HIROs is modulated - and in particular enhanced - at the magnetic fields where the PIRO condition is also met. This interplay was first observed by \citeauthor{Zhang2008b}~\cite{Zhang2008b} in a quantum well with high electron density.

The oscillatory component of the electron-phonon interaction is described theoretically within the framework of \citeauthor{Dmitriev2010}~\cite{Dmitriev2010}, who classify the HIRO-PIRO interaction into three regimes based on the ratio of the electron drift velocity $v_\mathrm{D}$ to the phonon sound velocity $s$: the subsonic regime ($\varepsilon_{\mathrm{dc}} < \varepsilon_{\mathrm{ph}}$, equivalently $v_\mathrm{D} < s$), a transition at the sound barrier ($v_\mathrm{D} = s$), and the supersonic regime ($v_{D} > s$). The theory predicts a characteristic phase shift in the oscillatory part of the electron-phonon interaction at the sound barrier. Recently, \citeauthor{Wang2026}~\cite{Wang2026} studied the development of PIROs under strong dc currents in the supersonic regime in an ultra-high-mobility GaAs quantum well, and confirmed the phase shift at the sound barrier experimentally.

An important aspect of this framework concerns the low-temperature behavior of the electron-phonon scattering rates~\cite{Dmitriev2010,Wang2026}. At temperatures $T < \hbar\omega_\mathrm{s}/k_\mathrm{B}$, the occupation of phonon modes with energy $\hbar\omega_\mathrm{s}$ is exponentially small with a corresponding suppression of electron-phonon scattering. However, spontaneous phonon emission depends on the electron occupation at the initial energies, which may be large in electric fields even for negligible thermal absorption.

In the present work, we study the subsonic regime in a GaAs/AlGaAs single-heterointerface two-dimensional electron gas (2DEG) — a system that differs from the quantum wells used in previous observations of the HIRO-PIRO interplay~\cite{Zhang2008b,Wang2026}. Under equilibrium conditions, our samples exhibit only two weak PIRO maxima up to temperatures of 6.6\,K. Despite these unfavorable conditions, we demonstrate that the dc current required to generate HIROs heats the electrons sufficiently to activate phonon emission. Through the double resonance with HIROs, the electron-phonon coupling becomes observable up to PIRO order $\varepsilon_{\mathrm{ph}} = 8$. We extract a sound velocity of $s = 3.4$\,km/s, identifying transverse acoustic phonons as the relevant mode, and provide independent experimental evidence for the phase shift at the sound barrier mentioned above.

In addition to a pristine reference sample, we investigate these effects in systems containing artificial scatterers in the shape of Swiss crosses, which form two-dimensional Lorentz gases (2DLGs). As reported in our companion paper~\cite{Bartels2026b}, these retroreflective obstacles can enhance the HIRO amplitude by an order of magnitude through an increased backscattering rate. Since the phonon-induced modulation is superimposed on the HIRO signal, this amplitude enhancement directly improves the visibility of the electron-phonon coupling. The Lorentz gases thus serve a dual purpose: they extend the observation of phonon modulation for higher electron temperatures while simultaneously providing a controlled test of whether artificial elastic scatterers affect the intrinsic electron-phonon interaction.

\section{Experimental Setup}\label{SecSampleExpSetup}

The measurements are performed on a GaAs/Al$_{0.315}$Ga$_{0.685}$As heterostructure with a 2DEG located 150\,nm below the surface at the single heterointerface. At liquid helium temperatures, the electron density is $n_\mathrm{e} = 2.64 \times 10^{15}$\,m$^{-2}$ and the mobility $\mu \approx 650$\,m$^2$/Vs, corresponding to a Fermi velocity of $v_\mathrm{F} \approx 2.22 \times 10^5$\,m/s. The Hall bar of width $W=100\,\upmu$m is patterned into the 2DEG by optical lithography. Alloyed Au/AuGe Ohmic contacts provide electrical access, and voltage probes of $10\,\upmu$m width with $100\,\upmu$m spacing enable four-terminal measurements.

In addition to a pristine reference sample, four samples from the same wafer contain randomly distributed Swiss cross-shaped obstacles defined by electron beam lithography and reactive ion etching. Each obstacle consists of two perpendicular rectangles of lithographic size $300\,\text{nm} \times 1850\,\text{nm}$, with an effective electronic size of $450\,\text{nm} \times 2000\,\text{nm}$ after accounting for a lateral depletion length of $\approx 75$\,nm~\cite{Schluck2015}. A scheme of a single obstacle is shown in Fig.~\ref{Fig_Mobility}. The dimensionless obstacle density $n^\star$ (fractional area coverage) takes values of $n^\star = 0.0098$, 0.0196, 0.0393, and 0.0785. A detailed characterization of the obstacle influence on the HIRO amplitude is presented in Ref.~\cite{Bartels2026b}.

Transport measurements are carried out in a $^3$He/$^4$He dilution refrigerator at bath temperatures between 40\,mK and 6.6\,K. The equilibrium electron temperature at the lowest bath temperature is approximately 100\,mK. Throughout, $T \equiv T_\mathrm{bath}$ denotes the bath temperature, while the electron temperature $T_\mathrm{e}\geq T$ depends on the dc current. An ac current of 500\,nA at 17.7\,Hz and a superimposed dc current (HF2LI, Zurich Instruments) are applied. The differential longitudinal resistivity $r_{xx}(I_{\mathrm{dc}}, B) \equiv V_{xx,\mathrm{ac}}(I_{\mathrm{dc}}, B)/I_{\mathrm{ac}}$ is recorded via lock-in amplifiers (SR 830 DSP, Stanford Research Systems). The dc current is swept from $-160\,\upmu$A to $+160\,\upmu$A in steps of $\approx 40$\,nA at magnetic fields up to 200\,mT with a step size of 1\,mT (at 40\,mK and 900\,mK) or 2.5\,mT (at 3.6\,K). Supplementary measurements at 1.5\,K, 2.2\,K, 2.8\,K, 5.6\,K, and 6.6\,K were performed at selected magnetic fields.

\section{Results and Discussion}\label{Sec_Results}

\subsection{Electron-phonon coupling under equilibrium conditions}\label{Sec_eph_equilibrium}

\begin{figure}[b]
\includegraphics[scale=1.0]{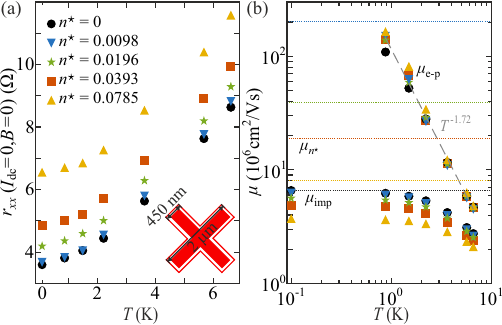}
	\caption{(a) Zero-field, zero-current resistivity $r_{xx}(B=0, I_{\mathrm{dc}}=0)$ versus temperature for all samples. Additionally a single Swiss-cross obstacle is shown, where the filled area corresponds to the lithographic size ($300\,\text{nm} \times 1850\,\text{nm}$ per arm) and the outer contour indicates the estimated electronic size, assuming a lateral depletion length of $\approx75\,\text{nm}$. (b) Mobility $\mu$ with individual contributions: impurity-limited mobility $\mu_\text{imp}$, obstacle-limited mobility $\mu_{n^\star}$, and electron-phonon mobility $\mu_{e\text{-}p} \propto T^{-1.72}$. The electron-phonon contribution is independent of $n^\star$.}
	\label{Fig_Mobility}
\end{figure}

Before discussing the nonequilibrium regime, we characterize the electron-phonon coupling under equilibrium conditions. Figure~\ref{Fig_Mobility}(a) shows the zero-field resistivity $r_{xx}(B=0, I_{\mathrm{dc}}=0)$ versus temperature for all samples. The mobility [Fig.~\ref{Fig_Mobility}(b)], calculated as $\mu = 1/(en_\mathrm{e} r_{xx})$, can be decomposed via Matthiessen's rule:
\begin{equation}
    \mu^{-1}(T) = \mu_\mathrm{{imp}}^{-1} + \mu_{n^\star}^{-1} + \mu_{e\text{-}p}^{-1}(T), \label{Eq_Matthiessen}
\end{equation}
where $\mu_\text{imp}$ is the temperature-independent impurity-limited mobility, $\mu_{n^\star}$ accounts for scattering at the artificial obstacles, and $\mu_{e\text{-}p}(T)$ characterizes the electron-phonon interaction. The temperature-dependent component $\mu_{e\text{-}p}(T)$ is independent of the artificial scatterers for $T\ge2$\,K, with only minor deviations at lower temperatures. From a linear fit in the double-logarithmic plot between 0.9\,K and 6.6\,K, we determine $\mu_{e\text{-}p}(T)\propto T^{-1.72}$. This exponent is close to the $-1.8$ reported for a high-density GaAs quantum well where PIROs up to 8th order were observed without dc current~\cite{Hatke2009}. However, in our system this dependence holds over the entire measured temperature range, whereas in ultra-high-mobility samples a transition to a steeper $T^{-5}$ dependence characteristic for the deep Bloch-Gr\"uneisen regime is typically observed below 3\,K~\cite{Hatke2009}. The observed exponent thus reflects the crossover between the high-temperature limit $\mu_\mathrm{e\text{-}p} \propto T^{-1}$ and the Bloch-Gr\"uneisen regime~\cite{Hatke2009,Stormer1990,Mendez1984}.

Two observations are central for the subsequent discussion. First of all, the mobility $\mu_{e\text{-}p}$ determined by electron-phonon scattering is independent of $n^\star$ for $T \geq 2$\,K, demonstrating that the artificial obstacles do not modify the intrinsic electron-phonon coupling. This is expected considering that the obstacle scattering is elastic, and the phonon spectrum is a bulk property of the GaAs crystal. Second, all measurements are performed below the Bloch-Gr\"uneisen temperature $T_\mathrm{BG} = 2\hbar k_\mathrm{F} s / k_\mathrm{B} = 6.69$\,K (with $s = 3.4$\,km/s, see Sec.~\ref{Sec_eph_subsonic}), where electron heating effects are dominant~\cite{Raichev2017}.

\subsection{Hall field induced resistance oscillations and electron heating}\label{Sec_Heating}

\begin{figure}
\includegraphics[scale=1.0]{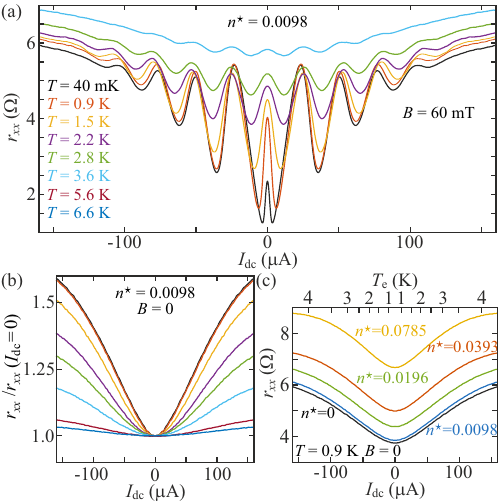}
	\caption{(a) Symmetrized resistivity $r_{xx}(I_\mathrm{dc})$ at $B = 60$\,mT for $n^\star = 0.0098$ at bath temperatures from 40\,mK to 3.6\,K, showing HIROs up to fifth order superimposed on the heating background. (b) Normalized zero-field resistivity $r_{xx}(B=0, I_\mathrm{dc})/r_{xx}(B=0, I_\mathrm{dc}=0)$ versus dc current for $n^\star = 0.0098$ for different bath temperatures up to 6.6\,K [legend in (a)]. (c) Zero-field resistivity $r_{xx}(B=0, I_\mathrm{dc})$ for all samples at $T = 0.9$\,K. The upper axis shows the estimated electron temperature $T_\mathrm{e}$, obtained from the equilibrium calibration $r_{xx}(I_\mathrm{dc}=0, B=0, T)$ shown in Fig.~\ref{Fig_Mobility}(a) via Eq.~\eqref{Eq_Tcalib}.}
	\label{Fig_HIROs}
\end{figure}

HIROs are measured in fixed perpendicular magnetic fields $B$ as a function of the dc current $I_{\mathrm{dc}}$. Figure~\ref{Fig_HIROs}(a) shows the symmetrized resistivity $r_{xx}(I_\mathrm{dc})$ at $B = 60$\,mT for several bath temperatures. HIROs up to fifth order ($\varepsilon_{\mathrm{dc}} = 5$) are superimposed on a monotonic heating background, following the resonance condition
\begin{equation}
\varepsilon_{\mathrm{dc}} = \gamma \sqrt{\frac{2\pi}{n_\mathrm{e}}} \frac{m^\star}{We^2} \frac{I_{\mathrm{dc}}}{B}, \label{Eq_edc}
\end{equation}
with $\gamma \approx 2$~\cite{Vavilov2007} and $m^\star = 0.067\,m_\mathrm{e}$. Throughout this work, $\varepsilon_{\mathrm{dc}}$ and $\varepsilon_{\mathrm{ph}}$ are treated as positive quantities, since only positive magnetic fields are applied and the data are symmetrized with respect to $I_{\mathrm{dc}}$, so that only $|\varepsilon_{\mathrm{dc}}|$ carries physical meaning. Because of the dominant role of electron heating, the data at 40\,mK and 900\,mK are nearly identical in the current range of interest (see Supplemental Material~\cite{SupplementBartels2026c}). The HIROs are characterized in Ref.~\cite{Bartels2026b}. Most important for the scope of the present study, the 2DLGs exhibit an enhanced HIRO amplitude within a specific regime due to the significantly increased ratio $\tau_\mathrm{D}/\tau_\pi$, where $\tau_\mathrm{D}$ is the Drude scattering time and $\tau_\pi$ the backscattering time~\cite{Bartels2026b}.

The dc current that generates HIROs simultaneously heats the electron system far above the bath temperature. This heating is directly visible in the increase of the zero-field resistivity $r_{xx}(B=0, I_{\mathrm{dc}})$ with increasing $I_{\mathrm{dc}}$ shown in Fig.~\ref{Fig_HIROs}(b)-(c). In Fig.~\ref{Fig_HIROs}(b), the normalized zero-field resistivity is shown for $n^\star = 0.0098$ at various bath temperatures. The relative increase with $I_{\mathrm{dc}}$ decreases monotonically for higher bath temperatures, consistent with electron heating in the Bloch-Gr\"uneisen regime~\cite{Raichev2017}: at higher $T_\text{bath}$, the electron system is already closer to the phonon-dominated transport regime, and additional current-induced heating produces a smaller relative change. The zero-field resistivity $r_{xx}(B=0, I_{\mathrm{dc}})$ for all $n^\star$ at $T=0.9$\,K is shown in Fig.~\ref{Fig_HIROs}(c). In the range $|I_{\mathrm{dc}}|< 60\,\upmu$A, the resistivity increases parabolically with $I_{\mathrm{dc}}$; at higher currents, the curve flattens out.

To quantify the electron heating, we use the equilibrium resistivity $r_{xx}(B=0, I_{\mathrm{dc}}=0, T)$ shown in Fig.~\ref{Fig_Mobility}(a) as thermometer, using the assumption that the resistivity at a given electron temperature $T_\mathrm{e}$ is the same regardless of whether $T_\mathrm{e}$ was established by thermal equilibrium with the bath or by current-induced heating. This assumption is consistent with the Bloch-Gr\"uneisen nonlinearity framework of \citeauthor{Raichev2017}~\cite{Raichev2017}, who showed that the nonlinear resistivity increase in this regime can be explained by the energy balance between current induced heating and phonon cooling of the electron system. Since the electron-phonon mobility $\mu_{e\text{-}p}$ is independent of $n^\star$ (Sec.~\ref{Sec_eph_equilibrium}), the temperature-dependent contribution to the resistivity is universal across all samples. We exploit this by fitting the equilibrium data to
\begin{equation}
    r_{xx}(T_\mathrm{e}) = r_{xx,0} + \alpha\,(T_\mathrm{e}^{1.72} - T_0^{1.72}), \label{Eq_Tcalib}
\end{equation}
where $r_{xx,0}$ is a sample-specific offset that absorbs the temperature-independent impurity and obstacle scattering, $T_0 = 0.9$\,K is the reference bath temperature, and $\alpha$ is a global coefficient determined from a simultaneous fit to all five samples. The electron temperature at finite $I_\mathrm{dc}$ is then obtained by inverting Eq.~\eqref{Eq_Tcalib} using the measured zero-field resistivity $r_{xx}(B=0, I_\mathrm{dc})$. The resulting $T_\mathrm{e}(I_\mathrm{dc})$ is approximately independent of $n^\star$ and indicated in the upper axis of Fig.~\ref{Fig_HIROs}(c).

At the highest applied currents ($|I_\mathrm{dc}| \approx 160\,\upmu$A), the electron temperature reaches $T_\mathrm{e} \gtrsim 4$\,K, well into the regime where emission of acoustic phonons becomes significant. The lattice temperature $T_\mathrm{L}$, in contrast, is expected to remain close to the bath temperature $T_\text{bath}$. This separation of $T_\mathrm{e}$ and $T_\mathrm{L}$ has important consequences for the microscopic mechanism by which the dc current activates the electron-phonon coupling, as discussed in the following section.

\subsection{Phonon-induced amplitude modulation in the subsonic regime}\label{Sec_eph_subsonic}

\begin{figure}
\includegraphics[scale=1.0]{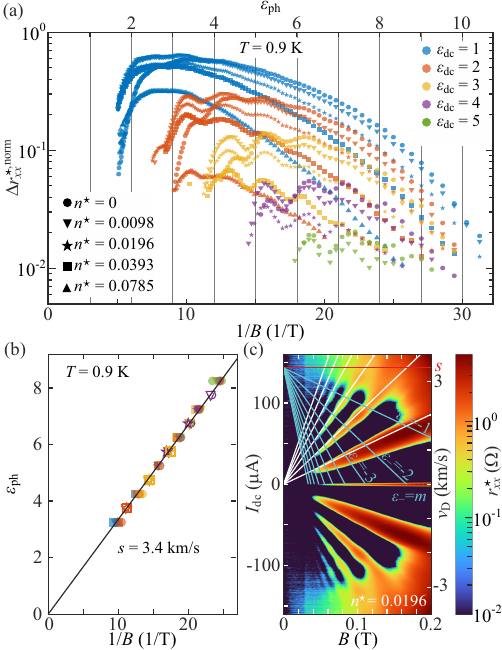}
	\caption{(a) Normalized HIRO amplitude $\Delta r_{xx}^{\star,\mathrm{norm}}$ (half the peak-to-peak value) versus $1/B$ for all samples and HIRO orders at $T = 0.9$\,K. The upper axis shows the PIRO order $\varepsilon_{\mathrm{ph}}$ for $s = 3.4$\,km/s. (b) Positions of amplitude modulation maxima (filled symbols) and minima (open symbols), plotted at $\varepsilon_{\mathrm{ph}} = k + 1/4$ ($\varepsilon_{\mathrm{ph}} = k - 1/4$) with positive integer $k$, versus $1/B$ for all samples and $\varepsilon_{\mathrm{dc}}$. The legends in (a) apply to (b) as well. The linear fit through the origin yields $s = 3.4$\,km/s. (c) Background-corrected resistivity $r_{xx}^\star$ versus $B$ and $I_\mathrm{dc}$ at 0.9\,K for $n^\star=0.0196$ (note the logarithmic color scale). The right axis shows the corresponding drift velocity $v_\mathrm{D}$. In the upper half, the white lines represent the expected HIRO positions for $\gamma = 2$, and the blue lines indicate integer values of $\varepsilon_{\mathrm{-}}$ with $m=n - k \in \mathbb{Z}, m<0$. The horizontal red line marks the sound barrier $v_\mathrm{D} = s$. As a result of the symmetrization procedure, the data are symmetric about $I_\mathrm{dc} = 0$.}
	\label{Fig_PIRO}
\end{figure}

The activated electron-phonon coupling manifests itself as an oscillatory modulation of the HIRO amplitude. To isolate it, we subtract the heating background introduced in Sec.~\ref{Sec_Heating} from $r_{xx}$, yielding the background-corrected resistivity $r_{xx}^\star$. Fig.~\ref{Fig_PIRO}(a) shows the normalized HIRO amplitude $\Delta r_{xx}^{\star,\mathrm{norm}} = \Delta r_{xx}^\star / r_{xx}(I_\mathrm{dc}=0,B) $, i.e., half the peak-to-peak value of the oscillation, versus $1/B$ for all HIRO orders at $T = 0.9$\,K for all samples. Without electron-phonon interactions and electron heating, the amplitude should decrease linearly in this representation. The observed oscillatory modulation is a clear signature of electron-phonon coupling, most pronounced for $\varepsilon_{\mathrm{dc}} \geq 2$, which is consistent with the increased electron heating at higher $\varepsilon_{\mathrm{dc}}$. At $\varepsilon_{\mathrm{dc}} = 2$ and $B = 60$\,mT, the required current of $\approx 60\,\upmu$A already increases $T_\mathrm{e}$ above 2\,K, where phonon emission becomes appreciable. At larger magnetic fields (smaller $1/B$), the HIRO amplitude is enhanced by the obstacles, an effect discussed in detail elsewhere~\cite{Bartels2026b}. Since the oscillations persist over the widest range of $1/B$ in this regime, the phonon-induced modulation imprinted on their envelope is also extracted more reliably and precisely there.

At $T_L \approx 0.9$\,K, the thermal occupation of acoustic phonon modes at the relevant energy $\hbar\omega_\mathrm{s}$ is exponentially small, and phonon absorption is strongly suppressed. However, the elevated electron temperature broadens the Fermi distribution, populating states above $E_\mathrm{F}$ and leaving partial vacancies in Landau levels below $E_\mathrm{F}$. This enables spontaneous phonon emission by electrons scattering into these partially emptied states~\cite{Wang2026}, which is the dominant electron-phonon coupling mechanism in this regime (i.e., in the subsonic regime at low bath temperatures). At higher bath temperatures, e.g. 3.6\,K, the equilibrium electron temperature alone is already sufficient to drive phonon emission without additional dc-induced heating, so that modulations of the HIRO amplitude appear even for $\varepsilon_{\mathrm{dc}}=1$. In addition, the now thermally populated modes open phonon absorption as a further channel~\cite{SupplementBartels2026c}.

The oscillatory part of the electron-phonon scattering rate $\Gamma_\mathrm{ph}^{\mathrm{(osc)}}$ in the subsonic regime ($\varepsilon_{\mathrm{dc}}<\varepsilon_{\mathrm{ph}}$) for low temperatures $\hbar\omega_\mathrm{c}/k_\mathrm{B} \lesssim T \lesssim T_\mathrm{BG}$ is given by~\cite{Dmitriev2010}
\begin{equation}
\begin{split}
&\Gamma_{\mathrm{ph}}^{(\mathrm{osc})} = \frac{4\delta^2 g^2 \omega_\mathrm{c} \varepsilon_\mathrm{ph}}{\pi^2 \varepsilon_\mathrm{dc}^{3/2}} \exp{(-2|\varepsilon_{-}|\hbar\omega_\mathrm{c}/k_\mathrm{B}T)}\times \\
&\left[ \sqrt{|\varepsilon_{-}|}\cos{(2\pi\varepsilon_{-})} + \sqrt{\varepsilon_{+}}e^{-2\varepsilon_{\mathrm{dc}}\hbar \omega_\mathrm{c}/k_\mathrm{B}T}\sin{(2\pi\varepsilon_{+})} \right]
\end{split} \label{OscRate}
\end{equation}
where $\delta = \exp{(-\pi/\omega_\mathrm{c}\tau_\mathrm{qH})}$ is the Dingle factor with the quantum scattering time $\tau_\mathrm{qH}$, $g$ is the dimensionless electron-phonon coupling constant and $\varepsilon_\pm = \varepsilon_{\mathrm{dc}} \pm \varepsilon_{\mathrm{ph}}$. The two oscillatory terms correspond to the two phonon processes introduced above: the $\varepsilon_{+}$-term represents absorption (energy change $\hbar \omega_\mathrm{H} + \hbar \omega_\mathrm{s}$), weighted by the thermal phonon occupation, and the $\varepsilon_{-}$-term represents emission ($\hbar \omega_\mathrm{H} - \hbar \omega_\mathrm{s}$). Eq.~\eqref{OscRate} was derived for thermal equilibrium ($T_\mathrm{L} = T_\mathrm{e}$), where it predicts an exponential freeze-out of the modulation at low $T$. In our case, the absorption ($\varepsilon_{+}$) term is frozen at $T_\mathrm{L} = 0.9$\,K, whereas the emission ($\varepsilon_{-}$) term survives because it is governed by $T_\mathrm{e}$ (see above). Retaining only the emission term and assuming therefore $T\rightarrow T_\mathrm{e}$, Eq.~\eqref{OscRate} reduces to
\begin{equation}
\Gamma_{\mathrm{ph}}^{(\mathrm{osc})} \propto e^{{-2|\varepsilon_{-}|\hbar\omega_\mathrm{c}/k_\mathrm{B}T_\mathrm{e}}}
 \sqrt{\frac{|\varepsilon_{-}|}{\varepsilon_\mathrm{dc}^3}}\cos{(2\pi\varepsilon_{-})}. \label{OscRate2}
\end{equation}
The oscillatory resistivity is obtained from $r_{xx}^{(\mathrm{osc})}/\rho_\mathrm{D} = \tau_\mathrm{D} \partial/\partial\varepsilon_{\mathrm{dc}}[\varepsilon_{\mathrm{dc}}\,\Gamma_{\mathrm{ph}}^{(\mathrm{osc})}]$, where $\rho_\mathrm{D} = r_{xx}(B = 0, I_\mathrm{dc} = 0)$ is the Drude resistivity [Fig.~\ref{Fig_Mobility}(a)]. Keeping only the leading term gives
\begin{equation}
r_{xx}^{(\mathrm{osc})}/\rho_\mathrm{D} =- \frac{8\delta^2g^2 \omega_\mathrm{s} \tau_\mathrm{D}}{\pi \sqrt{\varepsilon_\mathrm{dc}}}\sqrt{|\varepsilon_{-}|} e^{{-2|\varepsilon_{-}|\hbar\omega_\mathrm{c}/k_\mathrm{B}T_\mathrm{e}}}
\sin(2\pi\varepsilon_-). \label{EqRxxOscPart}
\end{equation}
In the range of validity of Eq. \eqref{OscRate} and the assumptions made in Eq. \eqref{OscRate2}, the other terms of $r_{xx}^{(\mathrm{osc})}$ [$\propto \cos{(2\pi\varepsilon_{-})}$] are negligible and lead only to minor changes in amplitude and phase~\cite{SupplementBartels2026c}. At a HIRO resonance, $\varepsilon_\mathrm{dc} = n \in \mathbb{N}$, Eq.~\eqref{EqRxxOscPart} thus reduces to $r_{xx}^{\mathrm{(osc)}} \propto \sin{(2\pi\varepsilon_\mathrm{ph})}$ and predicts amplitude maxima at $\varepsilon_{\mathrm{ph}} = k + 1/4$ and minima at $\varepsilon_{\mathrm{ph}} = k - 1/4$ ($k \in \mathbb{N}$).

Crucially, this establishes an equidistant spacing of the extrema ($\Delta\varepsilon_{\mathrm{ph}} = 1$), i.e., the oscillation period is constant~\cite{SupplementBartels2026c}. This relation enables us to extract the sound velocity. Figure~\ref{Fig_PIRO}(b) shows the maxima (minima) positions, plotted as $\varepsilon_{\mathrm{ph}} = k + 1/4$ ($\varepsilon_{\mathrm{ph}} = k - 1/4$) versus $1/B$ for all samples and HIRO orders. Since the modulation positions of all samples fall onto a common line, a single fit including all obstacle densities is justified, consistent with the phonon mode being an intrinsic bulk property of the host crystal unaffected by the artificial scatterers. The linear fit through the origin yields $s = 3.4$\,km/s, in excellent agreement with the bulk transverse acoustic phonon velocity $s_t^{||}=3.37$\,km/s~\cite{Cottam1973,Bartels2026cNote}. Previously reported values extracted from PIROs in GaAs range between 3--6\,km/s~\cite{Zudov2001,Zhang2008b,Lei2008b,Hatke2009b,Bykov2010,Hatke2011, Raichev2017}, depending on the heterostructure, which can affect the coupling to different phonon branches and the relative weight of bulk versus interface modes.

Using the calculated sound velocity, the upper axis in Fig.~\ref{Fig_PIRO}(a) shows that the electron-phonon coupling is resolved up to $\varepsilon_{\mathrm{ph}} = 8$. This is remarkable considering that under equilibrium conditions, only two weak PIRO maxima are detectable, even at 6.6\,K~\cite{SupplementBartels2026c}. This does not stem from an enhanced coupling constant $g$. Rather, electron heating lifts the exponential suppression of phonon emission, but the resulting modulation remains small, between 0.02\,$\Omega$ and 0.2\,$\Omega$ in the evaluated range. It becomes detectable only because it is imprinted on HIRO amplitudes determined with an uncertainty below this value; a less precise amplitude determination would obscure the modulation. In equilibrium, by contrast, the PIROs are masked by the non-trivial $\rho_{xx}(B)$ background.

The data from the Lorentz gas samples confirm the $n^\star$-independence of the phonon modulation positions. The obstacles modify the elastic scattering landscape and enhance the HIRO amplitude at high electric fields~\cite{Bartels2026b}, but they leave the phonon modulation period unchanged. An influence on the modulation amplitude, and thus on the electron-phonon coupling constant, cannot be resolved: the amplitude is small and depends on several sample- and temperature-dependent quantities, such as $T_\mathrm{e}$ and $\tau_\mathrm{qH}(T)$, whose uncertainties exceed possible differences between the samples. We therefore determine $g$ as an average over all samples. The benefit of the obstacles for the phonon analysis is thus indirect: by enhancing the HIRO amplitude, they can increase the signal on which the phonon-induced modulation is imprinted, which may render electron-phonon effects accessible in samples with intrinsically weak HIROs.

Using Eq.~\eqref{EqRxxOscPart} and the estimated electron temperature $T_\mathrm{e}(I_\mathrm{dc})$ with the corresponding $\tau_{\mathrm{qH}}(T)$ from Ref.~\cite{Bartels2026b}, we obtain $g^2 \approx 0.0049 \pm 0.0025$, averaged over all samples, $\varepsilon_{\mathrm{dc}} \in \{3,4\}$ and $\varepsilon_{\mathrm{ph}} \in \{4,5,6\}$. This is about a factor of 3 larger than the value reported in Ref.~\cite{Wang2026} at 3.9\,K bath temperature, which may further contribute to the visibility of the modulation up to $\varepsilon_{\mathrm{ph}} = 8$ in our sample.

The full resonance structure is summarized in Fig.~\ref{Fig_PIRO}(c). In the $B-I_\mathrm{dc}$ plane, the HIRO maxima (white) appear as straight lines $I_\mathrm{dc} \propto B$. The phonon-induced modulation maxima follow $\varepsilon_\mathrm{-} = m - 1/4$ ($m = n-k \in \mathbb{Z}, m <0$), i.e., they lie close to, but importantly not at, the integer values of $\varepsilon_-$ marked by the blue lines. The horizontal red line marks the sound barrier ($v_\mathrm{D} = s$), separating the subsonic and supersonic regimes.

The persistence of the modulation up to $\varepsilon_\mathrm{ph} = 8$ at $T_\mathrm{L} = 0.9$\,K, together with the quarter-period offset of its maxima relative to $\varepsilon_\mathrm{ph}$, confirms that the coupling is driven by spontaneous phonon emission at the elevated electron temperature $T_\mathrm{e}$. Equations~\eqref{OscRate2} and \eqref{EqRxxOscPart} capture this through a leading-order approximation. Eq.~\eqref{OscRate} is a single-temperature expression in which the phonon occupation and the electronic Fermi factors are linked by detailed balance at a common $T$. In our evaluation we relax this condition by assigning the lattice temperature $T_\mathrm{L}$ to the phonon occupation and the electron temperature $T_\mathrm{e}$ to the Fermi factors, which breaks the equilibrium relation between emission and absorption. While numerical approaches based on balance equations, see Ref.~\cite{Lei2008b}, determine $T_\mathrm{e}$ self-consistently and capture the current-induced enhancement of the oscillations, they do not provide analytical expressions. Nevertheless, the excellent agreement between the observed oscillation peak positions and the predictions of Ref.~\cite{Dmitriev2010} indicates that the single-temperature description captures the fundamental resonance condition correctly, i.e., it is dictated by the energy matching between the Landau level spacing and the phonon energy. The development of a comprehensive analytical quantum kinetic theory that incorporates separate electron and lattice temperatures thus remains an open challenge~\cite{Wang2026}.

\subsection{Phase shift at the sound barrier}\label{Sec_soundbarrier}

\begin{figure}
\includegraphics[scale=1.0]{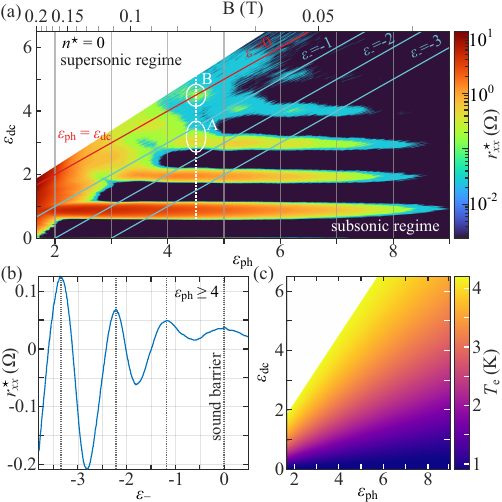}
	\caption{(a) Background-corrected resistivity $r_{xx}^\star$ versus $\varepsilon_{\mathrm{dc}}$ and $\varepsilon_{\mathrm{ph}}$ for $n^\star = 0$ at 0.9\,K on a logarithmic color scale. Only positive values of $\varepsilon_{\mathrm{dc}}$ are shown due to the symmetrization procedure. The upper axis indicates the corresponding $B$ in 10\,mT steps. The diagonal blue lines mark integer values of $\varepsilon_- = \varepsilon_{\mathrm{dc}} - \varepsilon_{\mathrm{ph}}$, and the red diagonal line marks the sound barrier ($\varepsilon_- = 0$). The white dashed line indicates the direction along which $r_{xx}^\star$ is evaluated as a function of $\varepsilon_-$ in panel (b). The circles A and B highlight two modulation maxima: for A, the maximum lies below the blue diagonal (at $\varepsilon_- = m - 1/4$), whereas at the sound barrier the phase shift moves the maximum onto the diagonal ($\varepsilon_- = 0$, circle B). (b) $r_{xx}^\star$ versus $\varepsilon_-$, averaged over $\varepsilon_{\mathrm{ph}} \geq 4$. The phase shift at $\varepsilon_- = 0$ is evident. (c) Estimated electron temperature versus $\varepsilon_{\mathrm{dc}}$ and $\varepsilon_{\mathrm{ph}}$, calculated from $r_{xx}(B = 0, I_{\mathrm{dc}})$ using the equilibrium calibration [Eq.~\eqref{Eq_Tcalib}].}
	\label{Fig_edceph}
\end{figure}

In Fig.~\ref{Fig_edceph}(a), the differential resistivity of the pristine sample is represented as a function of $\varepsilon_{\mathrm{dc}}$ and $\varepsilon_{\mathrm{ph}}$. As throughout this work, $\varepsilon_{\mathrm{dc}}$ takes only positive values, since the data have been symmetrized with respect to $I_{\mathrm{dc}}$. The red diagonal marks the sound barrier ($\varepsilon_- =0$), separating the subsonic ($\varepsilon_{\mathrm{dc}} < \varepsilon_{\mathrm{ph}}$) and supersonic ($\varepsilon_{\mathrm{dc}} > \varepsilon_{\mathrm{ph}}$) regimes while the blue diagonals mark integer values of $\varepsilon_-$. The red and blue lines mark the same $\varepsilon_-$ values as in Fig.~\ref{Fig_PIRO}(c), providing a direct correspondence between the two otherwise non-trivially related representations. The amplitude modulation is visible for $\varepsilon_{\mathrm{dc}} \geq 2$. The estimated electron temperature [Fig.~\ref{Fig_edceph}(c)], obtained from the calibration established in Sec.~\ref{Sec_Heating}, confirms that the modulation is confined to regions where $T_\mathrm{e}$ exceeds $\approx 2$\,K.

According to Eq.~\eqref{EqRxxOscPart}, the phase of the electron-phonon interaction is governed by the combined parameter $\varepsilon_{-}$. As the drift velocity $v_\mathrm{D}$ approaches the sound velocity $s$, Ref.~\cite{Dmitriev2010} predicts that the oscillation undergoes a characteristic $\pi/2$ phase shift. In the subsonic regime its maxima lie at $\varepsilon_{-} = m - 1/4$. Across the sound barrier they shift by a quarter period and are at integer values of $\varepsilon_{-}$~\cite{SupplementBartels2026c}. This becomes directly visible in Fig.~\ref{Fig_edceph}(a), where the white dashed line marks, as an example, a vertical cut at constant $\varepsilon_{\mathrm{ph}}$ (i.e.\ constant $B$). Along this cut, the phase of the modulation is determined by $\varepsilon_-$ alone, while its amplitude still depends on $\varepsilon_{\mathrm{dc}}$ and $T_\mathrm{e}$. The white circles illustrate the phase shift: maximum A, in the subsonic regime, lies at $\varepsilon_- \approx -1.25$ under the blue diagonal of $\varepsilon_{-} = -1$, whereas maximum B, at the sound barrier, has moved onto the diagonal $\varepsilon_- = 0$.

Figure~\ref{Fig_edceph}(b) shows such cuts averaged over all $\varepsilon_{\mathrm{ph}} \geq 4$, which confirms the phase shift discussed above. This finding independently corroborates the recent observation by \citeauthor{Wang2026}~\cite{Wang2026}, who studied the development of PIROs in the supersonic regime. The phase shift occurs everywhere along the sound barrier, which corresponds to a fixed dc current $I_\mathrm{dc} \approx 140\,\upmu$A ($v_\mathrm{D} = s$) and thus to elevated $T_\mathrm{e}$ at every order. Near this large current the PIRO modulation lies on a steep heating background whose subtraction is reliable only at the correspondingly lower magnetic fields (higher $\varepsilon_\mathrm{ph}$). Therefore, in our data it can be resolved only for $\varepsilon_\mathrm{ph} \geq 4$. There, in turn, the HIRO amplitude of the Lorentz gases is suppressed, since the enhancement of $\tau_\mathrm{D}/\tau_\pi$ decreases at elevated $T_\mathrm{e}$ and the increased quantum scattering together with Hall-field inhomogeneities damp the oscillations~\cite{Bartels2026b}, so that only the pristine sample with its large Dingle factor retains enough amplitude for the shift to be resolved. This does not contradict the enhanced visibility of the modulation in the 2DLGs, which occurs at intermediate fields and for lower orders, well inside the subsonic regime.

Taken together with the analysis of the modulation period and amplitude in Sec.~\ref{Sec_eph_subsonic} and the finding of Ref.~\cite{Wang2026}, the present results show that the phonon-induced modulation agrees with the theoretical framework of Ref.~\cite{Dmitriev2010} on both sides of the sound barrier. This verification is subject to a general limitation: at the low lattice temperatures examined here, the modulation becomes observable only because the dc current heats the electron system well above the lattice temperature, whereas the theory assumes a common temperature of electrons and phonons.

\section{Summary}\label{SecSummary}

We have reported that the electron heating by the dc current used to generate Hall field induced resistance oscillations causes hot electrons to emit spontaneously acoustic phonons in a GaAs/AlGaAs single-heterointerface 2DEG, even at bath temperatures as low as 40\,mK. This phonon emission scatters the electrons into the partial vacancies that the elevated electron temperature opens in the Landau levels below $E_\mathrm{F}$, while the lattice remains close to the bath temperature and thermal phonon absorption is negligible. For higher bath temperatures, phonon absorption becomes an additional scattering channel. The resulting double resonance between HIROs and PIROs renders the phonon-induced modulation observable up to PIRO order $\varepsilon_{\mathrm{ph}} = 8$, although only two weak PIRO maxima are detectable under equilibrium conditions. The active phonon mode is identified as the transverse acoustic phonon with a sound velocity of $s = 3.4$\,km/s. The characteristic phase of the oscillatory modulation in the subsonic regime is confirmed and consistent with the theoretical predictions of Ref.~\cite{Dmitriev2010}. At the sound barrier, we observe the predicted phase shift, independently confirming the recent experimental verification by \citeauthor{Wang2026}~\cite{Wang2026}. The convergence of these complementary findings provides a comprehensive validation of the theoretical framework for the oscillatory electron-phonon interaction in strongly tilted Landau levels. A model incorporating separate electron and lattice temperatures, as called for by both our observations as well as by those of Ref.~\cite{Wang2026}, remains to be developed.

The measurements on the Lorentz gas samples confirm that electron-phonon coupling is an intrinsic property of the GaAs host crystal, unaffected by the artificial obstacles that modify the elastic scattering landscape. These obstacles enhance the HIRO amplitude through retroreflective scattering~\cite{Bartels2026b}, which extends the range of visibility of the double resonances discussed here to larger magnetic fields / dc currents.

\section{Acknowledgments}
The authors acknowledge financial support by the Heinrich Heine University Düsseldorf.

\section{Data Availability}
The data supporting the findings of this article are openly available~\cite{DAS_Bartels2026c}.


\begin{thebibliography}{62}%
\makeatletter
\providecommand \@ifxundefined [1]{%
 \@ifx{#1\undefined}
}%
\providecommand \@ifnum [1]{%
 \ifnum #1\expandafter \@firstoftwo
 \else \expandafter \@secondoftwo
 \fi
}%
\providecommand \@ifx [1]{%
 \ifx #1\expandafter \@firstoftwo
 \else \expandafter \@secondoftwo
 \fi
}%
\providecommand \natexlab [1]{#1}%
\providecommand \enquote  [1]{``#1''}%
\providecommand \bibnamefont  [1]{#1}%
\providecommand \bibfnamefont [1]{#1}%
\providecommand \citenamefont [1]{#1}%
\providecommand \href@noop [0]{\@secondoftwo}%
\providecommand \href [0]{\begingroup \@sanitize@url \@href}%
\providecommand \@href[1]{\@@startlink{#1}\@@href}%
\providecommand \@@href[1]{\endgroup#1\@@endlink}%
\providecommand \@sanitize@url [0]{\catcode `\\12\catcode `\$12\catcode `\&12\catcode `\#12\catcode `\^12\catcode `\_12\catcode `\%12\relax}%
\providecommand \@@startlink[1]{}%
\providecommand \@@endlink[0]{}%
\providecommand \url  [0]{\begingroup\@sanitize@url \@url }%
\providecommand \@url [1]{\endgroup\@href {#1}{\urlprefix }}%
\providecommand \urlprefix  [0]{URL }%
\providecommand \Eprint [0]{\href }%
\providecommand \doibase [0]{https://doi.org/}%
\providecommand \selectlanguage [0]{\@gobble}%
\providecommand \bibinfo  [0]{\@secondoftwo}%
\providecommand \bibfield  [0]{\@secondoftwo}%
\providecommand \translation [1]{[#1]}%
\providecommand \BibitemOpen [0]{}%
\providecommand \bibitemStop [0]{}%
\providecommand \bibitemNoStop [0]{.\EOS\space}%
\providecommand \EOS [0]{\spacefactor3000\relax}%
\providecommand \BibitemShut  [1]{\csname bibitem#1\endcsname}%
\let\auto@bib@innerbib\@empty
\bibitem [{\citenamefont {Gurevich}\ and\ \citenamefont {Firsov}(1961)}]{Gurevich1961}%
  \BibitemOpen
  \bibfield  {author} {\bibinfo {author} {\bibfnamefont {V.~L.}\ \bibnamefont {Gurevich}}\ and\ \bibinfo {author} {\bibfnamefont {Y.~A.}\ \bibnamefont {Firsov}},\ }\bibfield  {title} {\bibinfo {title} {{On the theory of the electrical conductivity of semiconductors in a magnetic field. I}},\ }\href {https://www.jetp.ras.ru/cgi-bin/dn/e_013_01_0137.pdf} {\bibfield  {journal} {\bibinfo  {journal} {Sov. Phys. JETP}\ }\textbf {\bibinfo {volume} {13}},\ \bibinfo {pages} {137} (\bibinfo {year} {1961})}\BibitemShut {NoStop}%
\bibitem [{\citenamefont {Yang}\ \emph {et~al.}(2002{\natexlab{a}})\citenamefont {Yang}, \citenamefont {Zudov}, \citenamefont {Zhang}, \citenamefont {Du}, \citenamefont {Simmons},\ and\ \citenamefont {Reno}}]{Yang2002b}%
  \BibitemOpen
  \bibfield  {author} {\bibinfo {author} {\bibfnamefont {C.~L.}\ \bibnamefont {Yang}}, \bibinfo {author} {\bibfnamefont {M.~A.}\ \bibnamefont {Zudov}}, \bibinfo {author} {\bibfnamefont {J.}~\bibnamefont {Zhang}}, \bibinfo {author} {\bibfnamefont {R.-R.}\ \bibnamefont {Du}}, \bibinfo {author} {\bibfnamefont {J.~A.}\ \bibnamefont {Simmons}},\ and\ \bibinfo {author} {\bibfnamefont {J.~L.}\ \bibnamefont {Reno}},\ }\bibfield  {title} {\bibinfo {title} {Magnetophonon resonance of two-dimensional electrons by leaky interface-acoustic phonons},\ }\href {https://doi.org/10.1016/S1386-9477(01)00313-7} {\bibfield  {journal} {\bibinfo  {journal} {Physica E-low-dimensional Systems \& Nanostructures}\ }\textbf {\bibinfo {volume} {12}},\ \bibinfo {pages} {443} (\bibinfo {year} {2002}{\natexlab{a}})}\BibitemShut {NoStop}%
\bibitem [{\citenamefont {Bykov}\ \emph {et~al.}(2005{\natexlab{a}})\citenamefont {Bykov}, \citenamefont {Kalagin},\ and\ \citenamefont {Bakarov}}]{Bykov2005}%
  \BibitemOpen
  \bibfield  {author} {\bibinfo {author} {\bibfnamefont {A.~A.}\ \bibnamefont {Bykov}}, \bibinfo {author} {\bibfnamefont {A.~K.}\ \bibnamefont {Kalagin}},\ and\ \bibinfo {author} {\bibfnamefont {A.~K.}\ \bibnamefont {Bakarov}},\ }\bibfield  {title} {\bibinfo {title} {{Magnetophonon resonance in a GaAs quantum well with AlAs/GaAs superlattice barriers at high filling factors}},\ }\href {https://doi.org/10.1134/1.1996762} {\bibfield  {journal} {\bibinfo  {journal} {JETP Lett.}\ }\textbf {\bibinfo {volume} {81}},\ \bibinfo {pages} {523} (\bibinfo {year} {2005}{\natexlab{a}})}\BibitemShut {NoStop}%
\bibitem [{\citenamefont {Zhang}\ \emph {et~al.}(2004)\citenamefont {Zhang}, \citenamefont {Lyo}, \citenamefont {Du}, \citenamefont {Simmons},\ and\ \citenamefont {Reno}}]{Zhang2004b}%
  \BibitemOpen
  \bibfield  {author} {\bibinfo {author} {\bibfnamefont {J.}~\bibnamefont {Zhang}}, \bibinfo {author} {\bibfnamefont {S.~K.}\ \bibnamefont {Lyo}}, \bibinfo {author} {\bibfnamefont {R.~R.}\ \bibnamefont {Du}}, \bibinfo {author} {\bibfnamefont {J.~A.}\ \bibnamefont {Simmons}},\ and\ \bibinfo {author} {\bibfnamefont {J.~L.}\ \bibnamefont {Reno}},\ }\bibfield  {title} {\bibinfo {title} {{Oscillatory Magnetothermopower and Resonant Phonon Drag in a High-Mobility 2D Electron Gas}},\ }\href {https://doi.org/10.1103/PhysRevLett.92.156802} {\bibfield  {journal} {\bibinfo  {journal} {Phys. Rev. Lett.}\ }\textbf {\bibinfo {volume} {92}},\ \bibinfo {pages} {156802} (\bibinfo {year} {2004})}\BibitemShut {NoStop}%
\bibitem [{\citenamefont {Hatke}\ \emph {et~al.}(2009{\natexlab{a}})\citenamefont {Hatke}, \citenamefont {Zudov}, \citenamefont {Pfeiffer},\ and\ \citenamefont {West}}]{Hatke2009}%
  \BibitemOpen
  \bibfield  {author} {\bibinfo {author} {\bibfnamefont {A.~T.}\ \bibnamefont {Hatke}}, \bibinfo {author} {\bibfnamefont {M.~A.}\ \bibnamefont {Zudov}}, \bibinfo {author} {\bibfnamefont {L.~N.}\ \bibnamefont {Pfeiffer}},\ and\ \bibinfo {author} {\bibfnamefont {K.~W.}\ \bibnamefont {West}},\ }\bibfield  {title} {\bibinfo {title} {{Phonon-Induced Resistance Oscillations in 2D Systems with a Very High Electron Mobility}},\ }\href {https://doi.org/10.1103/PhysRevLett.102.086808} {\bibfield  {journal} {\bibinfo  {journal} {Phys. Rev. Lett.}\ }\textbf {\bibinfo {volume} {102}},\ \bibinfo {pages} {086808} (\bibinfo {year} {2009}{\natexlab{a}})}\BibitemShut {NoStop}%
\bibitem [{\citenamefont {Bykov}\ and\ \citenamefont {Goran}(2009)}]{Bykov2009}%
  \BibitemOpen
  \bibfield  {author} {\bibinfo {author} {\bibfnamefont {A.~A.}\ \bibnamefont {Bykov}}\ and\ \bibinfo {author} {\bibfnamefont {A.~V.}\ \bibnamefont {Goran}},\ }\bibfield  {title} {\bibinfo {title} {{Temperature dependence of magnetophonon resistance oscillations in GaAs/AlAs heterostructures at high filling factors}},\ }\href {https://doi.org/10.1134/S0021364009200065} {\bibfield  {journal} {\bibinfo  {journal} {JETP Lett.}\ }\textbf {\bibinfo {volume} {90}},\ \bibinfo {pages} {578} (\bibinfo {year} {2009})}\BibitemShut {NoStop}%
\bibitem [{\citenamefont {Bykov}\ \emph {et~al.}(2010)\citenamefont {Bykov}, \citenamefont {Goran},\ and\ \citenamefont {Vitkalov}}]{Bykov2010}%
  \BibitemOpen
  \bibfield  {author} {\bibinfo {author} {\bibfnamefont {A.~A.}\ \bibnamefont {Bykov}}, \bibinfo {author} {\bibfnamefont {A.~V.}\ \bibnamefont {Goran}},\ and\ \bibinfo {author} {\bibfnamefont {S.~A.}\ \bibnamefont {Vitkalov}},\ }\bibfield  {title} {\bibinfo {title} {{Interference of magneto-intersubband and phonon-induced resistance oscillations in single GaAs quantum wells with two populated subbands}},\ }\href {https://doi.org/10.1103/PhysRevB.81.155322} {\bibfield  {journal} {\bibinfo  {journal} {Phys. Rev. B}\ }\textbf {\bibinfo {volume} {81}},\ \bibinfo {pages} {155322} (\bibinfo {year} {2010})}\BibitemShut {NoStop}%
\bibitem [{\citenamefont {Hatke}\ \emph {et~al.}(2011{\natexlab{a}})\citenamefont {Hatke}, \citenamefont {Zudov}, \citenamefont {Pfeiffer},\ and\ \citenamefont {West}}]{Hatke2011}%
  \BibitemOpen
  \bibfield  {author} {\bibinfo {author} {\bibfnamefont {A.~T.}\ \bibnamefont {Hatke}}, \bibinfo {author} {\bibfnamefont {M.~A.}\ \bibnamefont {Zudov}}, \bibinfo {author} {\bibfnamefont {L.~N.}\ \bibnamefont {Pfeiffer}},\ and\ \bibinfo {author} {\bibfnamefont {K.~W.}\ \bibnamefont {West}},\ }\bibfield  {title} {\bibinfo {title} {Phase of phonon-induced resistance oscillations in a high-mobility two-dimensional electron gas},\ }\href {https://doi.org/10.1103/PhysRevB.84.121301} {\bibfield  {journal} {\bibinfo  {journal} {Phys. Rev. B}\ }\textbf {\bibinfo {volume} {84}},\ \bibinfo {pages} {121301} (\bibinfo {year} {2011}{\natexlab{a}})}\BibitemShut {NoStop}%
\bibitem [{\citenamefont {Kumaravadivel}\ \emph {et~al.}(2019)\citenamefont {Kumaravadivel}, \citenamefont {Greenaway}, \citenamefont {Perello}, \citenamefont {Berdyugin}, \citenamefont {Birkbeck}, \citenamefont {Wengraf}, \citenamefont {Liu}, \citenamefont {Edgar}, \citenamefont {Geim}, \citenamefont {Eaves},\ and\ \citenamefont {Krishna~Kumar}}]{Kumaravadivel2019}%
  \BibitemOpen
  \bibfield  {author} {\bibinfo {author} {\bibfnamefont {P.}~\bibnamefont {Kumaravadivel}}, \bibinfo {author} {\bibfnamefont {M.~T.}\ \bibnamefont {Greenaway}}, \bibinfo {author} {\bibfnamefont {D.}~\bibnamefont {Perello}}, \bibinfo {author} {\bibfnamefont {A.}~\bibnamefont {Berdyugin}}, \bibinfo {author} {\bibfnamefont {J.}~\bibnamefont {Birkbeck}}, \bibinfo {author} {\bibfnamefont {J.}~\bibnamefont {Wengraf}}, \bibinfo {author} {\bibfnamefont {S.}~\bibnamefont {Liu}}, \bibinfo {author} {\bibfnamefont {J.~H.}\ \bibnamefont {Edgar}}, \bibinfo {author} {\bibfnamefont {A.~K.}\ \bibnamefont {Geim}}, \bibinfo {author} {\bibfnamefont {L.}~\bibnamefont {Eaves}},\ and\ \bibinfo {author} {\bibfnamefont {R.}~\bibnamefont {Krishna~Kumar}},\ }\bibfield  {title} {\bibinfo {title} {Strong magnetophonon oscillations in extra-large graphene},\ }\href {https://doi.org/10.1038/s41467-019-11379-3} {\bibfield  {journal} {\bibinfo  {journal} {Nature Communications}\ }\textbf {\bibinfo {volume} {10}},\ \bibinfo {pages} {3334}
  (\bibinfo {year} {2019})}\BibitemShut {NoStop}%
\bibitem [{\citenamefont {Greenaway}\ \emph {et~al.}(2019)\citenamefont {Greenaway}, \citenamefont {Krishna~Kumar}, \citenamefont {Kumaravadivel}, \citenamefont {Geim},\ and\ \citenamefont {Eaves}}]{Greenaway2019}%
  \BibitemOpen
  \bibfield  {author} {\bibinfo {author} {\bibfnamefont {M.~T.}\ \bibnamefont {Greenaway}}, \bibinfo {author} {\bibfnamefont {R.}~\bibnamefont {Krishna~Kumar}}, \bibinfo {author} {\bibfnamefont {P.}~\bibnamefont {Kumaravadivel}}, \bibinfo {author} {\bibfnamefont {A.~K.}\ \bibnamefont {Geim}},\ and\ \bibinfo {author} {\bibfnamefont {L.}~\bibnamefont {Eaves}},\ }\bibfield  {title} {\bibinfo {title} {{Magnetophonon spectroscopy of Dirac fermion scattering by transverse and longitudinal acoustic phonons in graphene}},\ }\href {https://doi.org/10.1103/PhysRevB.100.155120} {\bibfield  {journal} {\bibinfo  {journal} {Phys. Rev. B}\ }\textbf {\bibinfo {volume} {100}},\ \bibinfo {pages} {155120} (\bibinfo {year} {2019})}\BibitemShut {NoStop}%
\bibitem [{\citenamefont {Tsui}\ \emph {et~al.}(1980)\citenamefont {Tsui}, \citenamefont {Englert}, \citenamefont {Cho},\ and\ \citenamefont {Gossard}}]{Tsui1980}%
  \BibitemOpen
  \bibfield  {author} {\bibinfo {author} {\bibfnamefont {D.~C.}\ \bibnamefont {Tsui}}, \bibinfo {author} {\bibfnamefont {T.}~\bibnamefont {Englert}}, \bibinfo {author} {\bibfnamefont {A.~Y.}\ \bibnamefont {Cho}},\ and\ \bibinfo {author} {\bibfnamefont {A.~C.}\ \bibnamefont {Gossard}},\ }\bibfield  {title} {\bibinfo {title} {{Observation of Magnetophonon Resonances in a Two-Dimensional Electronic System}},\ }\href {https://doi.org/10.1103/PhysRevLett.44.341} {\bibfield  {journal} {\bibinfo  {journal} {Phys. Rev. Lett.}\ }\textbf {\bibinfo {volume} {44}},\ \bibinfo {pages} {341} (\bibinfo {year} {1980})}\BibitemShut {NoStop}%
\bibitem [{\citenamefont {Englert}\ \emph {et~al.}(1982)\citenamefont {Englert}, \citenamefont {Tsui}, \citenamefont {Portal}, \citenamefont {Beerens},\ and\ \citenamefont {Gossard}}]{Englert1982}%
  \BibitemOpen
  \bibfield  {author} {\bibinfo {author} {\bibfnamefont {T.}~\bibnamefont {Englert}}, \bibinfo {author} {\bibfnamefont {D.~C.}\ \bibnamefont {Tsui}}, \bibinfo {author} {\bibfnamefont {J.}~\bibnamefont {Portal}}, \bibinfo {author} {\bibfnamefont {J.}~\bibnamefont {Beerens}},\ and\ \bibinfo {author} {\bibfnamefont {A.}~\bibnamefont {Gossard}},\ }\bibfield  {title} {\bibinfo {title} {{Magnetophonon resonances of the two-dimensional electron gas in GaAs/AlGaAs heterostructures}},\ }\href {https://doi.org/https://doi.org/10.1016/0038-1098(82)91109-7} {\bibfield  {journal} {\bibinfo  {journal} {Solid State Communications}\ }\textbf {\bibinfo {volume} {44}},\ \bibinfo {pages} {1301} (\bibinfo {year} {1982})}\BibitemShut {NoStop}%
\bibitem [{\citenamefont {Kido}\ \emph {et~al.}(1982)\citenamefont {Kido}, \citenamefont {Miura}, \citenamefont {Ohno},\ and\ \citenamefont {Sakaki}}]{Kido1982}%
  \BibitemOpen
  \bibfield  {author} {\bibinfo {author} {\bibfnamefont {G.}~\bibnamefont {Kido}}, \bibinfo {author} {\bibfnamefont {N.}~\bibnamefont {Miura}}, \bibinfo {author} {\bibfnamefont {H.}~\bibnamefont {Ohno}},\ and\ \bibinfo {author} {\bibfnamefont {H.}~\bibnamefont {Sakaki}},\ }\bibfield  {title} {\bibinfo {title} {{Magnetophonon Resonance in a Two-Dimensional Electron System in the GaAs–AlxGa1-xAs Heterojunction Interface}},\ }\href {https://doi.org/10.1143/JPSJ.51.2168} {\bibfield  {journal} {\bibinfo  {journal} {J. Phys. Soc. Jpn.}\ }\textbf {\bibinfo {volume} {51}},\ \bibinfo {pages} {121911} (\bibinfo {year} {1982})}\BibitemShut {NoStop}%
\bibitem [{\citenamefont {Zhang}\ \emph {et~al.}(2008)\citenamefont {Zhang}, \citenamefont {Zudov}, \citenamefont {Pfeiffer},\ and\ \citenamefont {West}}]{Zhang2008b}%
  \BibitemOpen
  \bibfield  {author} {\bibinfo {author} {\bibfnamefont {W.}~\bibnamefont {Zhang}}, \bibinfo {author} {\bibfnamefont {M.~A.}\ \bibnamefont {Zudov}}, \bibinfo {author} {\bibfnamefont {L.~N.}\ \bibnamefont {Pfeiffer}},\ and\ \bibinfo {author} {\bibfnamefont {K.~W.}\ \bibnamefont {West}},\ }\bibfield  {title} {\bibinfo {title} {{Resonant Phonon Scattering in Quantum Hall Systems Driven by dc Electric Fields}},\ }\href {https://doi.org/10.1103/PhysRevLett.100.036805} {\bibfield  {journal} {\bibinfo  {journal} {Phys. Rev. Lett.}\ }\textbf {\bibinfo {volume} {100}},\ \bibinfo {pages} {036805} (\bibinfo {year} {2008})}\BibitemShut {NoStop}%
\bibitem [{\citenamefont {Wang}\ \emph {et~al.}(2026)\citenamefont {Wang}, \citenamefont {Hilke}, \citenamefont {Fong}, \citenamefont {Austing}, \citenamefont {Studenikin}, \citenamefont {West},\ and\ \citenamefont {Pfeiffer}}]{Wang2026}%
  \BibitemOpen
  \bibfield  {author} {\bibinfo {author} {\bibfnamefont {Z.~T.}\ \bibnamefont {Wang}}, \bibinfo {author} {\bibfnamefont {M.}~\bibnamefont {Hilke}}, \bibinfo {author} {\bibfnamefont {N.}~\bibnamefont {Fong}}, \bibinfo {author} {\bibfnamefont {D.~G.}\ \bibnamefont {Austing}}, \bibinfo {author} {\bibfnamefont {S.~A.}\ \bibnamefont {Studenikin}}, \bibinfo {author} {\bibfnamefont {K.~W.}\ \bibnamefont {West}},\ and\ \bibinfo {author} {\bibfnamefont {L.~N.}\ \bibnamefont {Pfeiffer}},\ }\bibfield  {title} {\bibinfo {title} {{Resonant Magnetophonon Emission by Supersonic Electrons in Ultrahigh-Mobility Two-Dimensional Systems}},\ }\href {https://doi.org/10.1103/m1nb-j1h6} {\bibfield  {journal} {\bibinfo  {journal} {Phys. Rev. Lett.}\ }\textbf {\bibinfo {volume} {136}},\ \bibinfo {pages} {146302} (\bibinfo {year} {2026})}\BibitemShut {NoStop}%
\bibitem [{\citenamefont {Yang}\ \emph {et~al.}(2002{\natexlab{b}})\citenamefont {Yang}, \citenamefont {Zhang}, \citenamefont {Du}, \citenamefont {Simmons},\ and\ \citenamefont {Reno}}]{Yang2002}%
  \BibitemOpen
  \bibfield  {author} {\bibinfo {author} {\bibfnamefont {C.~L.}\ \bibnamefont {Yang}}, \bibinfo {author} {\bibfnamefont {J.}~\bibnamefont {Zhang}}, \bibinfo {author} {\bibfnamefont {R.~R.}\ \bibnamefont {Du}}, \bibinfo {author} {\bibfnamefont {J.~A.}\ \bibnamefont {Simmons}},\ and\ \bibinfo {author} {\bibfnamefont {J.~L.}\ \bibnamefont {Reno}},\ }\bibfield  {title} {\bibinfo {title} {{Zener Tunneling Between Landau Orbits in a High-Mobility Two-Dimensional Electron Gas}},\ }\href {https://doi.org/10.1103/PhysRevLett.89.076801} {\bibfield  {journal} {\bibinfo  {journal} {Phys. Rev. Lett.}\ }\textbf {\bibinfo {volume} {89}},\ \bibinfo {pages} {076801} (\bibinfo {year} {2002}{\natexlab{b}})}\BibitemShut {NoStop}%
\bibitem [{\citenamefont {Bykov}\ \emph {et~al.}(2005{\natexlab{b}})\citenamefont {Bykov}, \citenamefont {Zhang}, \citenamefont {Vitkalov}, \citenamefont {Kalagin},\ and\ \citenamefont {Bakarov}}]{Bykov2005b}%
  \BibitemOpen
  \bibfield  {author} {\bibinfo {author} {\bibfnamefont {A.~A.}\ \bibnamefont {Bykov}}, \bibinfo {author} {\bibfnamefont {J.}~\bibnamefont {Zhang}}, \bibinfo {author} {\bibfnamefont {S.}~\bibnamefont {Vitkalov}}, \bibinfo {author} {\bibfnamefont {A.~K.}\ \bibnamefont {Kalagin}},\ and\ \bibinfo {author} {\bibfnamefont {A.~K.}\ \bibnamefont {Bakarov}},\ }\bibfield  {title} {\bibinfo {title} {{Effect of dc and ac excitations on the longitudinal resistance of a two-dimensional electron gas in highly doped GaAs quantum wells}},\ }\href {https://doi.org/10.1103/PhysRevB.72.245307} {\bibfield  {journal} {\bibinfo  {journal} {Phys. Rev. B}\ }\textbf {\bibinfo {volume} {72}},\ \bibinfo {pages} {245307} (\bibinfo {year} {2005}{\natexlab{b}})}\BibitemShut {NoStop}%
\bibitem [{\citenamefont {Zhang}\ \emph {et~al.}(2007{\natexlab{a}})\citenamefont {Zhang}, \citenamefont {Chiang}, \citenamefont {Zudov}, \citenamefont {Pfeiffer},\ and\ \citenamefont {West}}]{Zhang2007}%
  \BibitemOpen
  \bibfield  {author} {\bibinfo {author} {\bibfnamefont {W.}~\bibnamefont {Zhang}}, \bibinfo {author} {\bibfnamefont {H.-S.}\ \bibnamefont {Chiang}}, \bibinfo {author} {\bibfnamefont {M.~A.}\ \bibnamefont {Zudov}}, \bibinfo {author} {\bibfnamefont {L.~N.}\ \bibnamefont {Pfeiffer}},\ and\ \bibinfo {author} {\bibfnamefont {K.~W.}\ \bibnamefont {West}},\ }\bibfield  {title} {\bibinfo {title} {{Magnetotransport in a two-dimensional electron system in dc electric fields}},\ }\href {https://doi.org/https://doi.org/10.1103/PhysRevB.75.041304} {\bibfield  {journal} {\bibinfo  {journal} {Phys. Rev. B}\ }\textbf {\bibinfo {volume} {75}},\ \bibinfo {pages} {041304(R)} (\bibinfo {year} {2007}{\natexlab{a}})}\BibitemShut {NoStop}%
\bibitem [{\citenamefont {Zhang}\ \emph {et~al.}(2007{\natexlab{b}})\citenamefont {Zhang}, \citenamefont {Vitkalov}, \citenamefont {Bykov}, \citenamefont {Kalagin},\ and\ \citenamefont {Bakarov}}]{Zhang2007b}%
  \BibitemOpen
  \bibfield  {author} {\bibinfo {author} {\bibfnamefont {J.-Q.}\ \bibnamefont {Zhang}}, \bibinfo {author} {\bibfnamefont {S.}~\bibnamefont {Vitkalov}}, \bibinfo {author} {\bibfnamefont {A.~A.}\ \bibnamefont {Bykov}}, \bibinfo {author} {\bibfnamefont {A.~K.}\ \bibnamefont {Kalagin}},\ and\ \bibinfo {author} {\bibfnamefont {A.~K.}\ \bibnamefont {Bakarov}},\ }\bibfield  {title} {\bibinfo {title} {Effect of a dc electric field on the longitudinal resistance of two-dimensional electrons in a magnetic field},\ }\href {https://doi.org/10.1103/PhysRevB.75.081305} {\bibfield  {journal} {\bibinfo  {journal} {Phys. Rev. B}\ }\textbf {\bibinfo {volume} {75}},\ \bibinfo {pages} {081305} (\bibinfo {year} {2007}{\natexlab{b}})}\BibitemShut {NoStop}%
\bibitem [{\citenamefont {Vavilov}\ \emph {et~al.}(2007)\citenamefont {Vavilov}, \citenamefont {Aleiner},\ and\ \citenamefont {Glazman}}]{Vavilov2007}%
  \BibitemOpen
  \bibfield  {author} {\bibinfo {author} {\bibfnamefont {M.~G.}\ \bibnamefont {Vavilov}}, \bibinfo {author} {\bibfnamefont {I.~L.}\ \bibnamefont {Aleiner}},\ and\ \bibinfo {author} {\bibfnamefont {L.~I.}\ \bibnamefont {Glazman}},\ }\bibfield  {title} {\bibinfo {title} {Nonlinear resistivity of a two-dimensional electron gas in a magnetic field},\ }\href {https://doi.org/10.1103/PhysRevB.76.115331} {\bibfield  {journal} {\bibinfo  {journal} {Phys. Rev. B}\ }\textbf {\bibinfo {volume} {76}},\ \bibinfo {pages} {115331} (\bibinfo {year} {2007})}\BibitemShut {NoStop}%
\bibitem [{\citenamefont {Lei}(2007)}]{Lei2007}%
  \BibitemOpen
  \bibfield  {author} {\bibinfo {author} {\bibfnamefont {X.~L.}\ \bibnamefont {Lei}},\ }\bibfield  {title} {\bibinfo {title} {Current-induced magnetoresistance oscillations in two-dimensional electron systems},\ }\href {https://doi.org/10.1063/1.2717572} {\bibfield  {journal} {\bibinfo  {journal} {Applied Physics Letters}\ }\textbf {\bibinfo {volume} {90}},\ \bibinfo {pages} {132119} (\bibinfo {year} {2007})}\BibitemShut {NoStop}%
\bibitem [{\citenamefont {Bykov}(2008)}]{Bykov2008}%
  \BibitemOpen
  \bibfield  {author} {\bibinfo {author} {\bibfnamefont {A.~A.}\ \bibnamefont {Bykov}},\ }\bibfield  {title} {\bibinfo {title} {{Zener tunneling between Landau levels in a double quantum well at high filling factors}},\ }\href {https://doi.org/10.1134/S0021364008180112} {\bibfield  {journal} {\bibinfo  {journal} {JETP Lett.}\ }\textbf {\bibinfo {volume} {88}},\ \bibinfo {pages} {394} (\bibinfo {year} {2008})}\BibitemShut {NoStop}%
\bibitem [{\citenamefont {Vitkalov}(2009)}]{Vitkalov2009}%
  \BibitemOpen
  \bibfield  {author} {\bibinfo {author} {\bibfnamefont {S.}~\bibnamefont {Vitkalov}},\ }\bibfield  {title} {\bibinfo {title} {{Nonlinear transport of 2D electrons in crossed electric and quantizing magnetic fields}},\ }\href {https://doi.org/10.1142/S0217979209054090} {\bibfield  {journal} {\bibinfo  {journal} {International Journal of Modern Physics B}\ }\textbf {\bibinfo {volume} {23}},\ \bibinfo {pages} {4727} (\bibinfo {year} {2009})}\BibitemShut {NoStop}%
\bibitem [{\citenamefont {Hatke}\ \emph {et~al.}(2009{\natexlab{b}})\citenamefont {Hatke}, \citenamefont {Zudov}, \citenamefont {Pfeiffer},\ and\ \citenamefont {West}}]{Hatke2009b}%
  \BibitemOpen
  \bibfield  {author} {\bibinfo {author} {\bibfnamefont {A.~T.}\ \bibnamefont {Hatke}}, \bibinfo {author} {\bibfnamefont {M.~A.}\ \bibnamefont {Zudov}}, \bibinfo {author} {\bibfnamefont {L.~N.}\ \bibnamefont {Pfeiffer}},\ and\ \bibinfo {author} {\bibfnamefont {K.~W.}\ \bibnamefont {West}},\ }\bibfield  {title} {\bibinfo {title} {{Role of electron-electron interactions in nonlinear transport in two-dimensional electron systems}},\ }\href {https://doi.org/10.1103/PhysRevB.79.161308} {\bibfield  {journal} {\bibinfo  {journal} {Phys. Rev. B}\ }\textbf {\bibinfo {volume} {79}},\ \bibinfo {pages} {161308} (\bibinfo {year} {2009}{\natexlab{b}})}\BibitemShut {NoStop}%
\bibitem [{\citenamefont {Hatke}\ \emph {et~al.}(2011{\natexlab{b}})\citenamefont {Hatke}, \citenamefont {Zudov}, \citenamefont {Pfeiffer},\ and\ \citenamefont {West}}]{Hatke2011b}%
  \BibitemOpen
  \bibfield  {author} {\bibinfo {author} {\bibfnamefont {A.~T.}\ \bibnamefont {Hatke}}, \bibinfo {author} {\bibfnamefont {M.~A.}\ \bibnamefont {Zudov}}, \bibinfo {author} {\bibfnamefont {L.~N.}\ \bibnamefont {Pfeiffer}},\ and\ \bibinfo {author} {\bibfnamefont {K.~W.}\ \bibnamefont {West}},\ }\bibfield  {title} {\bibinfo {title} {{Resistance oscillations induced by the Hall field in tilted magnetic fields}},\ }\href {https://doi.org/10.1103/PhysRevB.83.081301} {\bibfield  {journal} {\bibinfo  {journal} {Phys. Rev. B}\ }\textbf {\bibinfo {volume} {83}},\ \bibinfo {pages} {081301(R)} (\bibinfo {year} {2011}{\natexlab{b}})}\BibitemShut {NoStop}%
\bibitem [{\citenamefont {Bykov}\ \emph {et~al.}(2012)\citenamefont {Bykov}, \citenamefont {Dmitriev}, \citenamefont {Marchishin}, \citenamefont {Byrnes},\ and\ \citenamefont {Vitkalov}}]{Bykov2012}%
  \BibitemOpen
  \bibfield  {author} {\bibinfo {author} {\bibfnamefont {A.~A.}\ \bibnamefont {Bykov}}, \bibinfo {author} {\bibfnamefont {D.~V.}\ \bibnamefont {Dmitriev}}, \bibinfo {author} {\bibfnamefont {I.~V.}\ \bibnamefont {Marchishin}}, \bibinfo {author} {\bibfnamefont {S.}~\bibnamefont {Byrnes}},\ and\ \bibinfo {author} {\bibfnamefont {S.~A.}\ \bibnamefont {Vitkalov}},\ }\bibfield  {title} {\bibinfo {title} {{Zener tunneling between Landau orbits in two-dimensional electron Corbino rings}},\ }\href {https://doi.org/10.1063/1.4729590} {\bibfield  {journal} {\bibinfo  {journal} {Applied Physics Letters}\ }\textbf {\bibinfo {volume} {100}},\ \bibinfo {pages} {251602} (\bibinfo {year} {2012})}\BibitemShut {NoStop}%
\bibitem [{\citenamefont {Hatke}\ \emph {et~al.}(2012{\natexlab{a}})\citenamefont {Hatke}, \citenamefont {Zudov}, \citenamefont {Pfeiffer},\ and\ \citenamefont {West}}]{Hatke2012b}%
  \BibitemOpen
  \bibfield  {author} {\bibinfo {author} {\bibfnamefont {A.~T.}\ \bibnamefont {Hatke}}, \bibinfo {author} {\bibfnamefont {M.~A.}\ \bibnamefont {Zudov}}, \bibinfo {author} {\bibfnamefont {L.~N.}\ \bibnamefont {Pfeiffer}},\ and\ \bibinfo {author} {\bibfnamefont {K.~W.}\ \bibnamefont {West}},\ }\bibfield  {title} {\bibinfo {title} {{Nonlinear response in overlapping and separated Landau levels of GaAs quantum wells}},\ }\href {https://doi.org/10.1103/PhysRevB.86.081307} {\bibfield  {journal} {\bibinfo  {journal} {Phys. Rev. B}\ }\textbf {\bibinfo {volume} {86}},\ \bibinfo {pages} {081307(R)} (\bibinfo {year} {2012}{\natexlab{a}})}\BibitemShut {NoStop}%
\bibitem [{\citenamefont {Studenikin}\ \emph {et~al.}(2012)\citenamefont {Studenikin}, \citenamefont {Granger}, \citenamefont {Kam}, \citenamefont {Sachrajda}, \citenamefont {Wasilewski},\ and\ \citenamefont {Poole}}]{Studenikin2012}%
  \BibitemOpen
  \bibfield  {author} {\bibinfo {author} {\bibfnamefont {S.~A.}\ \bibnamefont {Studenikin}}, \bibinfo {author} {\bibfnamefont {G.}~\bibnamefont {Granger}}, \bibinfo {author} {\bibfnamefont {A.}~\bibnamefont {Kam}}, \bibinfo {author} {\bibfnamefont {A.~S.}\ \bibnamefont {Sachrajda}}, \bibinfo {author} {\bibfnamefont {Z.~R.}\ \bibnamefont {Wasilewski}},\ and\ \bibinfo {author} {\bibfnamefont {P.~J.}\ \bibnamefont {Poole}},\ }\bibfield  {title} {\bibinfo {title} {{Nonlinear magnetotransport phenomena in high-mobility two-dimensional electrons in InGaAs/InP and GaAs/AlGaAs}},\ }\href {https://doi.org/10.1103/PhysRevB.86.115309} {\bibfield  {journal} {\bibinfo  {journal} {Phys. Rev. B}\ }\textbf {\bibinfo {volume} {86}},\ \bibinfo {pages} {115309} (\bibinfo {year} {2012})}\BibitemShut {NoStop}%
\bibitem [{\citenamefont {Dmitriev}\ \emph {et~al.}(2012)\citenamefont {Dmitriev}, \citenamefont {Mirlin}, \citenamefont {Polyakov},\ and\ \citenamefont {Zudov}}]{Dmitriev2012}%
  \BibitemOpen
  \bibfield  {author} {\bibinfo {author} {\bibfnamefont {I.~A.}\ \bibnamefont {Dmitriev}}, \bibinfo {author} {\bibfnamefont {A.~D.}\ \bibnamefont {Mirlin}}, \bibinfo {author} {\bibfnamefont {D.~G.}\ \bibnamefont {Polyakov}},\ and\ \bibinfo {author} {\bibfnamefont {M.~A.}\ \bibnamefont {Zudov}},\ }\bibfield  {title} {\bibinfo {title} {{Nonequilibrium phenomena in high Landau levels}},\ }\href {https://doi.org/10.1103/RevModPhys.84.1709} {\bibfield  {journal} {\bibinfo  {journal} {Rev. Mod. Phys.}\ }\textbf {\bibinfo {volume} {84}},\ \bibinfo {pages} {1709} (\bibinfo {year} {2012})}\BibitemShut {NoStop}%
\bibitem [{\citenamefont {Shi}\ \emph {et~al.}(2014{\natexlab{a}})\citenamefont {Shi}, \citenamefont {Ebner},\ and\ \citenamefont {Zudov}}]{Shi2014c}%
  \BibitemOpen
  \bibfield  {author} {\bibinfo {author} {\bibfnamefont {Q.}~\bibnamefont {Shi}}, \bibinfo {author} {\bibfnamefont {Q.~A.}\ \bibnamefont {Ebner}},\ and\ \bibinfo {author} {\bibfnamefont {M.~A.}\ \bibnamefont {Zudov}},\ }\bibfield  {title} {\bibinfo {title} {{Hall field-induced resistance oscillations in a p-type Ge/SiGe quantum well}},\ }\href {https://doi.org/10.1103/PhysRevB.90.161301} {\bibfield  {journal} {\bibinfo  {journal} {Phys. Rev. B}\ }\textbf {\bibinfo {volume} {90}},\ \bibinfo {pages} {161301} (\bibinfo {year} {2014}{\natexlab{a}})}\BibitemShut {NoStop}%
\bibitem [{\citenamefont {Shi}\ \emph {et~al.}(2017)\citenamefont {Shi}, \citenamefont {Zudov}, \citenamefont {Falson}, \citenamefont {Kozuka}, \citenamefont {Tsukazaki}, \citenamefont {Kawasaki}, \citenamefont {von Klitzing},\ and\ \citenamefont {Smet}}]{Shi2017}%
  \BibitemOpen
  \bibfield  {author} {\bibinfo {author} {\bibfnamefont {Q.}~\bibnamefont {Shi}}, \bibinfo {author} {\bibfnamefont {M.~A.}\ \bibnamefont {Zudov}}, \bibinfo {author} {\bibfnamefont {J.}~\bibnamefont {Falson}}, \bibinfo {author} {\bibfnamefont {Y.}~\bibnamefont {Kozuka}}, \bibinfo {author} {\bibfnamefont {A.}~\bibnamefont {Tsukazaki}}, \bibinfo {author} {\bibfnamefont {M.}~\bibnamefont {Kawasaki}}, \bibinfo {author} {\bibfnamefont {K.}~\bibnamefont {von Klitzing}},\ and\ \bibinfo {author} {\bibfnamefont {J.}~\bibnamefont {Smet}},\ }\bibfield  {title} {\bibinfo {title} {{Hall field-induced resistance oscillations in MgZnO/ZnO heterostructures}},\ }\href {https://doi.org/10.1103/PhysRevB.95.041411} {\bibfield  {journal} {\bibinfo  {journal} {Phys. Rev. B}\ }\textbf {\bibinfo {volume} {95}},\ \bibinfo {pages} {041411} (\bibinfo {year} {2017})}\BibitemShut {NoStop}%
\bibitem [{\citenamefont {Zudov}\ \emph {et~al.}(2017)\citenamefont {Zudov}, \citenamefont {Dmitriev}, \citenamefont {Friess}, \citenamefont {Shi}, \citenamefont {Umansky}, \citenamefont {von Klitzing},\ and\ \citenamefont {Smet}}]{Zudov2017}%
  \BibitemOpen
  \bibfield  {author} {\bibinfo {author} {\bibfnamefont {M.~A.}\ \bibnamefont {Zudov}}, \bibinfo {author} {\bibfnamefont {I.~A.}\ \bibnamefont {Dmitriev}}, \bibinfo {author} {\bibfnamefont {B.}~\bibnamefont {Friess}}, \bibinfo {author} {\bibfnamefont {Q.}~\bibnamefont {Shi}}, \bibinfo {author} {\bibfnamefont {V.}~\bibnamefont {Umansky}}, \bibinfo {author} {\bibfnamefont {K.}~\bibnamefont {von Klitzing}},\ and\ \bibinfo {author} {\bibfnamefont {J.}~\bibnamefont {Smet}},\ }\bibfield  {title} {\bibinfo {title} {{Hall field-induced resistance oscillations in a tunable-density GaAs quantum well}},\ }\href {https://doi.org/10.1103/PhysRevB.96.121301} {\bibfield  {journal} {\bibinfo  {journal} {Phys. Rev. B}\ }\textbf {\bibinfo {volume} {96}},\ \bibinfo {pages} {121301} (\bibinfo {year} {2017})}\BibitemShut {NoStop}%
\bibitem [{\citenamefont {Greenaway}\ \emph {et~al.}(2021)\citenamefont {Greenaway}, \citenamefont {Kumaravadivel}, \citenamefont {Wengraf}, \citenamefont {Ponomarenko}, \citenamefont {Berdyugin}, \citenamefont {Li}, \citenamefont {Edgar}, \citenamefont {Kumar}, \citenamefont {Geim},\ and\ \citenamefont {Eaves}}]{Greenaway2021}%
  \BibitemOpen
  \bibfield  {author} {\bibinfo {author} {\bibfnamefont {M.~T.}\ \bibnamefont {Greenaway}}, \bibinfo {author} {\bibfnamefont {P.}~\bibnamefont {Kumaravadivel}}, \bibinfo {author} {\bibfnamefont {J.}~\bibnamefont {Wengraf}}, \bibinfo {author} {\bibfnamefont {L.~A.}\ \bibnamefont {Ponomarenko}}, \bibinfo {author} {\bibfnamefont {A.~I.}\ \bibnamefont {Berdyugin}}, \bibinfo {author} {\bibfnamefont {J.}~\bibnamefont {Li}}, \bibinfo {author} {\bibfnamefont {J.~H.}\ \bibnamefont {Edgar}}, \bibinfo {author} {\bibfnamefont {R.~K.}\ \bibnamefont {Kumar}}, \bibinfo {author} {\bibfnamefont {A.~K.}\ \bibnamefont {Geim}},\ and\ \bibinfo {author} {\bibfnamefont {L.}~\bibnamefont {Eaves}},\ }\bibfield  {title} {\bibinfo {title} {{Graphene’s non-equilibrium fermions reveal Doppler-shifted magnetophonon resonances accompanied by Mach supersonic and Landau velocity effects}},\ }\href {https://doi.org/10.1038/s41467-021-26663-4} {\bibfield  {journal} {\bibinfo  {journal} {Nature Communications}\ }\textbf {\bibinfo {volume}
  {12}},\ \bibinfo {pages} {6392} (\bibinfo {year} {2021})}\BibitemShut {NoStop}%
\bibitem [{\citenamefont {Wang}\ \emph {et~al.}(2023)\citenamefont {Wang}, \citenamefont {Hilke}, \citenamefont {Fong}, \citenamefont {Austing}, \citenamefont {Studenikin}, \citenamefont {West},\ and\ \citenamefont {Pfeiffer}}]{Wang2023}%
  \BibitemOpen
  \bibfield  {author} {\bibinfo {author} {\bibfnamefont {Z.~T.}\ \bibnamefont {Wang}}, \bibinfo {author} {\bibfnamefont {M.}~\bibnamefont {Hilke}}, \bibinfo {author} {\bibfnamefont {N.}~\bibnamefont {Fong}}, \bibinfo {author} {\bibfnamefont {D.~G.}\ \bibnamefont {Austing}}, \bibinfo {author} {\bibfnamefont {S.~A.}\ \bibnamefont {Studenikin}}, \bibinfo {author} {\bibfnamefont {K.~W.}\ \bibnamefont {West}},\ and\ \bibinfo {author} {\bibfnamefont {L.~N.}\ \bibnamefont {Pfeiffer}},\ }\bibfield  {title} {\bibinfo {title} {Nonlinear transport phenomena and current-induced hydrodynamics in ultrahigh mobility two-dimensional electron gas},\ }\href {https://doi.org/10.1103/PhysRevB.107.195406} {\bibfield  {journal} {\bibinfo  {journal} {Phys. Rev. B}\ }\textbf {\bibinfo {volume} {107}},\ \bibinfo {pages} {195406} (\bibinfo {year} {2023})}\BibitemShut {NoStop}%
\bibitem [{\citenamefont {Bartels}\ \emph {et~al.}(2025)\citenamefont {Bartels}, \citenamefont {Strobel}, \citenamefont {Horn-Cosfeld}, \citenamefont {Cerchez}, \citenamefont {Pierz}, \citenamefont {Schumacher}, \citenamefont {Mailly},\ and\ \citenamefont {Heinzel}}]{Bartels2025}%
  \BibitemOpen
  \bibfield  {author} {\bibinfo {author} {\bibfnamefont {F.}~\bibnamefont {Bartels}}, \bibinfo {author} {\bibfnamefont {J.}~\bibnamefont {Strobel}}, \bibinfo {author} {\bibfnamefont {B.}~\bibnamefont {Horn-Cosfeld}}, \bibinfo {author} {\bibfnamefont {M.}~\bibnamefont {Cerchez}}, \bibinfo {author} {\bibfnamefont {K.}~\bibnamefont {Pierz}}, \bibinfo {author} {\bibfnamefont {H.~W.}\ \bibnamefont {Schumacher}}, \bibinfo {author} {\bibfnamefont {D.}~\bibnamefont {Mailly}},\ and\ \bibinfo {author} {\bibfnamefont {T.}~\bibnamefont {Heinzel}},\ }\bibfield  {title} {\bibinfo {title} {{Hall field induced resistance oscillations by designed random obstacle arrays}},\ }\href {https://doi.org/10.1103/PhysRevB.111.165301} {\bibfield  {journal} {\bibinfo  {journal} {Phys. Rev. B}\ }\textbf {\bibinfo {volume} {111}},\ \bibinfo {pages} {165301} (\bibinfo {year} {2025})}\BibitemShut {NoStop}%
\bibitem [{\citenamefont {Bartels}\ \emph {et~al.}()\citenamefont {Bartels}, \citenamefont {Cerchez}, \citenamefont {Lange}, \citenamefont {Mailly}, \citenamefont {Pierz}, \citenamefont {Schumacher},\ and\ \citenamefont {Heinzel}}]{Bartels2026b}%
  \BibitemOpen
  \bibfield  {author} {\bibinfo {author} {\bibfnamefont {F.}~\bibnamefont {Bartels}}, \bibinfo {author} {\bibfnamefont {M.}~\bibnamefont {Cerchez}}, \bibinfo {author} {\bibfnamefont {J.}~\bibnamefont {Lange}}, \bibinfo {author} {\bibfnamefont {D.}~\bibnamefont {Mailly}}, \bibinfo {author} {\bibfnamefont {K.}~\bibnamefont {Pierz}}, \bibinfo {author} {\bibfnamefont {H.~W.}\ \bibnamefont {Schumacher}},\ and\ \bibinfo {author} {\bibfnamefont {T.}~\bibnamefont {Heinzel}},\ }\bibfield  {title} {\bibinfo {title} {{Amplification of Hall field induced resistance oscillations by retroreflective obstacles}},\ }\href@noop {} {\bibinfo  {journal} {to be published}\ }\BibitemShut {NoStop}%
\bibitem [{\citenamefont {Dmitriev}\ \emph {et~al.}(2010)\citenamefont {Dmitriev}, \citenamefont {Gellmann},\ and\ \citenamefont {Vavilov}}]{Dmitriev2010}%
  \BibitemOpen
\bibfield  {journal} {  }\bibfield  {author} {\bibinfo {author} {\bibfnamefont {I.~A.}\ \bibnamefont {Dmitriev}}, \bibinfo {author} {\bibfnamefont {R.}~\bibnamefont {Gellmann}},\ and\ \bibinfo {author} {\bibfnamefont {M.~G.}\ \bibnamefont {Vavilov}},\ }\bibfield  {title} {\bibinfo {title} {{Phonon-induced resistance oscillations of two-dimensional electron systems drifting with supersonic velocities}},\ }\href {https://doi.org/10.1103/PhysRevB.82.201311} {\bibfield  {journal} {\bibinfo  {journal} {Phys. Rev. B}\ }\textbf {\bibinfo {volume} {82}},\ \bibinfo {pages} {201311} (\bibinfo {year} {2010})}\BibitemShut {NoStop}%
\bibitem [{\citenamefont {Schluck}\ \emph {et~al.}(2015)\citenamefont {Schluck}, \citenamefont {Fasbender}, \citenamefont {Heinzel}, \citenamefont {Pierz}, \citenamefont {Schumacher}, \citenamefont {Kazazis},\ and\ \citenamefont {Gennser}}]{Schluck2015}%
  \BibitemOpen
  \bibfield  {author} {\bibinfo {author} {\bibfnamefont {J.}~\bibnamefont {Schluck}}, \bibinfo {author} {\bibfnamefont {S.}~\bibnamefont {Fasbender}}, \bibinfo {author} {\bibfnamefont {T.}~\bibnamefont {Heinzel}}, \bibinfo {author} {\bibfnamefont {K.}~\bibnamefont {Pierz}}, \bibinfo {author} {\bibfnamefont {H.~W.}\ \bibnamefont {Schumacher}}, \bibinfo {author} {\bibfnamefont {D.}~\bibnamefont {Kazazis}},\ and\ \bibinfo {author} {\bibfnamefont {U.}~\bibnamefont {Gennser}},\ }\bibfield  {title} {\bibinfo {title} {Commensurability resonances in two-dimensional magnetoelectric lateral superlattices},\ }\href {https://doi.org/10.1103/PhysRevB.91.195303} {\bibfield  {journal} {\bibinfo  {journal} {Phys. Rev. B}\ }\textbf {\bibinfo {volume} {91}},\ \bibinfo {pages} {195303} (\bibinfo {year} {2015})}\BibitemShut {NoStop}%
\bibitem [{\citenamefont {Stormer}\ \emph {et~al.}(1990)\citenamefont {Stormer}, \citenamefont {Pfeiffer}, \citenamefont {Baldwin},\ and\ \citenamefont {West}}]{Stormer1990}%
  \BibitemOpen
  \bibfield  {author} {\bibinfo {author} {\bibfnamefont {H.~L.}\ \bibnamefont {Stormer}}, \bibinfo {author} {\bibfnamefont {L.~N.}\ \bibnamefont {Pfeiffer}}, \bibinfo {author} {\bibfnamefont {K.~W.}\ \bibnamefont {Baldwin}},\ and\ \bibinfo {author} {\bibfnamefont {K.~W.}\ \bibnamefont {West}},\ }\bibfield  {title} {\bibinfo {title} {{Observation of a Bloch-Gr\"uneisen regime in two-dimensional electron transport}},\ }\href {https://doi.org/10.1103/PhysRevB.41.1278} {\bibfield  {journal} {\bibinfo  {journal} {Phys. Rev. B}\ }\textbf {\bibinfo {volume} {41}},\ \bibinfo {pages} {1278} (\bibinfo {year} {1990})}\BibitemShut {NoStop}%
\bibitem [{\citenamefont {Mendez}\ \emph {et~al.}(1984)\citenamefont {Mendez}, \citenamefont {Price},\ and\ \citenamefont {Heiblum}}]{Mendez1984}%
  \BibitemOpen
  \bibfield  {author} {\bibinfo {author} {\bibfnamefont {E.~E.}\ \bibnamefont {Mendez}}, \bibinfo {author} {\bibfnamefont {P.~J.}\ \bibnamefont {Price}},\ and\ \bibinfo {author} {\bibfnamefont {M.}~\bibnamefont {Heiblum}},\ }\bibfield  {title} {\bibinfo {title} {{Temperature dependence of the electron mobility in GaAs‐GaAlAs heterostructures}},\ }\href {https://doi.org/10.1063/1.95178} {\bibfield  {journal} {\bibinfo  {journal} {Applied Physics Letters}\ }\textbf {\bibinfo {volume} {45}},\ \bibinfo {pages} {294} (\bibinfo {year} {1984})}\BibitemShut {NoStop}%
\bibitem [{\citenamefont {Raichev}\ \emph {et~al.}(2017)\citenamefont {Raichev}, \citenamefont {Hatke}, \citenamefont {Zudov},\ and\ \citenamefont {Reno}}]{Raichev2017}%
  \BibitemOpen
  \bibfield  {author} {\bibinfo {author} {\bibfnamefont {O.~E.}\ \bibnamefont {Raichev}}, \bibinfo {author} {\bibfnamefont {A.~T.}\ \bibnamefont {Hatke}}, \bibinfo {author} {\bibfnamefont {M.~A.}\ \bibnamefont {Zudov}},\ and\ \bibinfo {author} {\bibfnamefont {J.~L.}\ \bibnamefont {Reno}},\ }\bibfield  {title} {\bibinfo {title} {{Bloch-Gr\"uneisen nonlinearity of electron transport in GaAs/AlGaAs heterostructures}},\ }\href {https://doi.org/10.1103/PhysRevB.96.081407} {\bibfield  {journal} {\bibinfo  {journal} {Phys. Rev. B}\ }\textbf {\bibinfo {volume} {96}},\ \bibinfo {pages} {081407} (\bibinfo {year} {2017})}\BibitemShut {NoStop}%
\bibitem [{Sup()}]{SupplementBartels2026c}%
  \BibitemOpen
  \href@noop {} {}\bibinfo {note} {See Supplemental Material at [URL will be inserted by publisher] for details of the data analysis and for further supporting measurements. Supplemental Material also includes Refs.~\cite{Shubnikov1930,Hatke2012,Mani2013,Bockhorn2011,Bockhorn2013,Shi2014a, Shi2014b, Bockhorn2014, Wang2016, Samaraweera2018, Schluck2018a, Horn2021, Wang2022, Bockhorn2024,Bartels2026}}\BibitemShut {NoStop}%
\bibitem [{\citenamefont {Cottam}\ and\ \citenamefont {Saunders}(1973)}]{Cottam1973}%
  \BibitemOpen
  \bibfield  {author} {\bibinfo {author} {\bibfnamefont {R.~I.}\ \bibnamefont {Cottam}}\ and\ \bibinfo {author} {\bibfnamefont {G.~A.}\ \bibnamefont {Saunders}},\ }\bibfield  {title} {\bibinfo {title} {The elastic constants of gaas from 2 k to 320 k},\ }\href {https://doi.org/10.1088/0022-3719/6/13/011} {\bibfield  {journal} {\bibinfo  {journal} {Journal of Physics C: Solid State Physics}\ }\textbf {\bibinfo {volume} {6}},\ \bibinfo {pages} {2105} (\bibinfo {year} {1973})}\BibitemShut {NoStop}%
\bibitem [{Bar()}]{Bartels2026cNote}%
  \BibitemOpen
  \href@noop {} {}\bibinfo {note} {$s_t^{\parallel} = 3.37$\,km/s denotes the low-temperature sound velocity of the bulk transverse acoustic mode polarized along [001] in GaAs.}\BibitemShut {Stop}%
\bibitem [{\citenamefont {Zudov}\ \emph {et~al.}(2001)\citenamefont {Zudov}, \citenamefont {Ponomarev}, \citenamefont {Efros}, \citenamefont {Du}, \citenamefont {Simmons},\ and\ \citenamefont {Reno}}]{Zudov2001}%
  \BibitemOpen
  \bibfield  {author} {\bibinfo {author} {\bibfnamefont {M.~A.}\ \bibnamefont {Zudov}}, \bibinfo {author} {\bibfnamefont {I.~V.}\ \bibnamefont {Ponomarev}}, \bibinfo {author} {\bibfnamefont {A.~L.}\ \bibnamefont {Efros}}, \bibinfo {author} {\bibfnamefont {R.~R.}\ \bibnamefont {Du}}, \bibinfo {author} {\bibfnamefont {J.~A.}\ \bibnamefont {Simmons}},\ and\ \bibinfo {author} {\bibfnamefont {J.~L.}\ \bibnamefont {Reno}},\ }\bibfield  {title} {\bibinfo {title} {{New Class of Magnetoresistance Oscillations: Interaction of a Two-Dimensional Electron Gas with Leaky Interface Phonons}},\ }\href {https://doi.org/10.1103/PhysRevLett.86.3614} {\bibfield  {journal} {\bibinfo  {journal} {Phys. Rev. Lett.}\ }\textbf {\bibinfo {volume} {86}},\ \bibinfo {pages} {3614} (\bibinfo {year} {2001})}\BibitemShut {NoStop}%
\bibitem [{\citenamefont {Lei}(2008)}]{Lei2008b}%
  \BibitemOpen
  \bibfield  {author} {\bibinfo {author} {\bibfnamefont {X.~L.}\ \bibnamefont {Lei}},\ }\bibfield  {title} {\bibinfo {title} {{Low temperature electron-phonon resonance in dc-current-biased two-dimensional electron systems}},\ }\href {https://doi.org/10.1103/PhysRevB.77.205309} {\bibfield  {journal} {\bibinfo  {journal} {Phys. Rev. B}\ }\textbf {\bibinfo {volume} {77}},\ \bibinfo {pages} {205309} (\bibinfo {year} {2008})}\BibitemShut {NoStop}%
\bibitem [{\citenamefont {Bartels}\ \emph {et~al.}(2026{\natexlab{a}})\citenamefont {Bartels}, \citenamefont {Cerchez}, \citenamefont {Lange}, \citenamefont {Mailly}, \citenamefont {Pierz}, \citenamefont {Schumacher},\ and\ \citenamefont {Heinzel}}]{DAS_Bartels2026c}%
  \BibitemOpen
  \bibfield  {author} {\bibinfo {author} {\bibfnamefont {F.}~\bibnamefont {Bartels}}, \bibinfo {author} {\bibfnamefont {M.}~\bibnamefont {Cerchez}}, \bibinfo {author} {\bibfnamefont {J.}~\bibnamefont {Lange}}, \bibinfo {author} {\bibfnamefont {D.}~\bibnamefont {Mailly}}, \bibinfo {author} {\bibfnamefont {K.}~\bibnamefont {Pierz}}, \bibinfo {author} {\bibfnamefont {H.~W.}\ \bibnamefont {Schumacher}},\ and\ \bibinfo {author} {\bibfnamefont {T.}~\bibnamefont {Heinzel}},\ }\bibfield  {title} {\bibinfo {title} {{Phonon emission due to dc-driven electron heating in a two-dimensional electron gas containing artificial scatterers [Data set]}},\ }\bibfield  {journal} {\bibinfo  {journal} {Zenodo}\ }\href {https://doi.org/10.5281/zenodo.20492290} {10.5281/zenodo.20492290} (\bibinfo {year} {2026}{\natexlab{a}})\BibitemShut {NoStop}%
\bibitem [{\citenamefont {Shubnikow}\ and\ \citenamefont {de~Haas}(1930)}]{Shubnikov1930}%
  \BibitemOpen
  \bibfield  {author} {\bibinfo {author} {\bibfnamefont {L.}~\bibnamefont {Shubnikow}}\ and\ \bibinfo {author} {\bibfnamefont {W.~J.}\ \bibnamefont {de~Haas}},\ }\bibfield  {title} {\bibinfo {title} {{A New Phenomenon in the Change of Resistance in a Magnetic Field of Single Crystals of Bismuth}},\ }\href {https://doi.org/10.1038/126500a0} {\bibfield  {journal} {\bibinfo  {journal} {Nature}\ }\textbf {\bibinfo {volume} {126}},\ \bibinfo {pages} {500} (\bibinfo {year} {1930})}\BibitemShut {NoStop}%
\bibitem [{\citenamefont {Hatke}\ \emph {et~al.}(2012{\natexlab{b}})\citenamefont {Hatke}, \citenamefont {Zudov}, \citenamefont {Reno}, \citenamefont {Pfeiffer},\ and\ \citenamefont {West}}]{Hatke2012}%
  \BibitemOpen
  \bibfield  {author} {\bibinfo {author} {\bibfnamefont {A.~T.}\ \bibnamefont {Hatke}}, \bibinfo {author} {\bibfnamefont {M.~A.}\ \bibnamefont {Zudov}}, \bibinfo {author} {\bibfnamefont {J.~L.}\ \bibnamefont {Reno}}, \bibinfo {author} {\bibfnamefont {L.~N.}\ \bibnamefont {Pfeiffer}},\ and\ \bibinfo {author} {\bibfnamefont {K.~W.}\ \bibnamefont {West}},\ }\bibfield  {title} {\bibinfo {title} {Giant negative magnetoresistance in high-mobility two-dimensional electron systems},\ }\href {https://doi.org/10.1103/PhysRevB.85.081304} {\bibfield  {journal} {\bibinfo  {journal} {Phys. Rev. B}\ }\textbf {\bibinfo {volume} {85}},\ \bibinfo {pages} {081304(R)} (\bibinfo {year} {2012}{\natexlab{b}})}\BibitemShut {NoStop}%
\bibitem [{\citenamefont {Mani}\ \emph {et~al.}(2013)\citenamefont {Mani}, \citenamefont {Kriisa},\ and\ \citenamefont {Wegscheider}}]{Mani2013}%
  \BibitemOpen
  \bibfield  {author} {\bibinfo {author} {\bibfnamefont {R.~G.}\ \bibnamefont {Mani}}, \bibinfo {author} {\bibfnamefont {A.}~\bibnamefont {Kriisa}},\ and\ \bibinfo {author} {\bibfnamefont {W.}~\bibnamefont {Wegscheider}},\ }\bibfield  {title} {\bibinfo {title} {{Size-dependent giant-magnetoresistance in millimeter scale GaAs/AlGaAs 2D electron devices}},\ }\href {https://doi.org/10.1038/srep02747} {\bibfield  {journal} {\bibinfo  {journal} {Scientific Reports}\ }\textbf {\bibinfo {volume} {3}},\ \bibinfo {pages} {02747} (\bibinfo {year} {2013})}\BibitemShut {NoStop}%
\bibitem [{\citenamefont {Bockhorn}\ \emph {et~al.}(2011)\citenamefont {Bockhorn}, \citenamefont {Barthold}, \citenamefont {Schuh}, \citenamefont {Wegscheider},\ and\ \citenamefont {Haug}}]{Bockhorn2011}%
  \BibitemOpen
  \bibfield  {author} {\bibinfo {author} {\bibfnamefont {L.}~\bibnamefont {Bockhorn}}, \bibinfo {author} {\bibfnamefont {P.}~\bibnamefont {Barthold}}, \bibinfo {author} {\bibfnamefont {D.}~\bibnamefont {Schuh}}, \bibinfo {author} {\bibfnamefont {W.}~\bibnamefont {Wegscheider}},\ and\ \bibinfo {author} {\bibfnamefont {R.~J.}\ \bibnamefont {Haug}},\ }\bibfield  {title} {\bibinfo {title} {Magnetoresistance in a high-mobility two-dimensional electron gas},\ }\href {https://doi.org/10.1103/PhysRevB.83.113301} {\bibfield  {journal} {\bibinfo  {journal} {Phys. Rev. B}\ }\textbf {\bibinfo {volume} {83}},\ \bibinfo {pages} {113301} (\bibinfo {year} {2011})}\BibitemShut {NoStop}%
\bibitem [{\citenamefont {Bockhorn}\ \emph {et~al.}(2013)\citenamefont {Bockhorn}, \citenamefont {Hodei}, \citenamefont {Schuh}, \citenamefont {Wegscheider},\ and\ \citenamefont {Haug}}]{Bockhorn2013}%
  \BibitemOpen
  \bibfield  {author} {\bibinfo {author} {\bibfnamefont {L.}~\bibnamefont {Bockhorn}}, \bibinfo {author} {\bibfnamefont {A.}~\bibnamefont {Hodei}}, \bibinfo {author} {\bibfnamefont {D.}~\bibnamefont {Schuh}}, \bibinfo {author} {\bibfnamefont {W.}~\bibnamefont {Wegscheider}},\ and\ \bibinfo {author} {\bibfnamefont {R.~J.}\ \bibnamefont {Haug}},\ }\bibfield  {title} {\bibinfo {title} {Magnetoresistance in a high mobility two-dimensional electron system as a function of sample geometry},\ }\href {https://doi.org/10.1088/1742-6596/456/1/012003} {\bibfield  {journal} {\bibinfo  {journal} {J. Phys. Conf. Ser.}\ }\textbf {\bibinfo {volume} {456}},\ \bibinfo {pages} {012003} (\bibinfo {year} {2013})}\BibitemShut {NoStop}%
\bibitem [{\citenamefont {Shi}\ \emph {et~al.}(2014{\natexlab{b}})\citenamefont {Shi}, \citenamefont {Martin}, \citenamefont {Ebner}, \citenamefont {Zudov}, \citenamefont {Pfeiffer},\ and\ \citenamefont {West}}]{Shi2014a}%
  \BibitemOpen
  \bibfield  {author} {\bibinfo {author} {\bibfnamefont {Q.}~\bibnamefont {Shi}}, \bibinfo {author} {\bibfnamefont {P.~D.}\ \bibnamefont {Martin}}, \bibinfo {author} {\bibfnamefont {Q.~A.}\ \bibnamefont {Ebner}}, \bibinfo {author} {\bibfnamefont {M.~A.}\ \bibnamefont {Zudov}}, \bibinfo {author} {\bibfnamefont {L.~N.}\ \bibnamefont {Pfeiffer}},\ and\ \bibinfo {author} {\bibfnamefont {K.~W.}\ \bibnamefont {West}},\ }\bibfield  {title} {\bibinfo {title} {Colossal negative magnetoresistance in a two-dimensional electron gas},\ }\href {https://doi.org/10.1103/PhysRevB.89.201301} {\bibfield  {journal} {\bibinfo  {journal} {Phys. Rev. B}\ }\textbf {\bibinfo {volume} {89}},\ \bibinfo {pages} {201301(R)} (\bibinfo {year} {2014}{\natexlab{b}})}\BibitemShut {NoStop}%
\bibitem [{\citenamefont {Shi}\ \emph {et~al.}(2014{\natexlab{c}})\citenamefont {Shi}, \citenamefont {Zudov}, \citenamefont {Pfeiffer},\ and\ \citenamefont {West}}]{Shi2014b}%
  \BibitemOpen
  \bibfield  {author} {\bibinfo {author} {\bibfnamefont {Q.}~\bibnamefont {Shi}}, \bibinfo {author} {\bibfnamefont {M.~A.}\ \bibnamefont {Zudov}}, \bibinfo {author} {\bibfnamefont {L.~N.}\ \bibnamefont {Pfeiffer}},\ and\ \bibinfo {author} {\bibfnamefont {K.~W.}\ \bibnamefont {West}},\ }\bibfield  {title} {\bibinfo {title} {{Nonlinear transport in two-dimensional electron gas exhibiting colossal negative magnetoresistance}},\ }\href {https://doi.org/10.1103/PhysRevB.90.201301} {\bibfield  {journal} {\bibinfo  {journal} {Phys. Rev. B}\ }\textbf {\bibinfo {volume} {90}},\ \bibinfo {pages} {201301(R)} (\bibinfo {year} {2014}{\natexlab{c}})}\BibitemShut {NoStop}%
\bibitem [{\citenamefont {Bockhorn}\ \emph {et~al.}(2014)\citenamefont {Bockhorn}, \citenamefont {Gornyi}, \citenamefont {Schuh}, \citenamefont {Reichl}, \citenamefont {Wegscheider},\ and\ \citenamefont {Haug}}]{Bockhorn2014}%
  \BibitemOpen
  \bibfield  {author} {\bibinfo {author} {\bibfnamefont {L.}~\bibnamefont {Bockhorn}}, \bibinfo {author} {\bibfnamefont {I.~V.}\ \bibnamefont {Gornyi}}, \bibinfo {author} {\bibfnamefont {D.}~\bibnamefont {Schuh}}, \bibinfo {author} {\bibfnamefont {C.}~\bibnamefont {Reichl}}, \bibinfo {author} {\bibfnamefont {W.}~\bibnamefont {Wegscheider}},\ and\ \bibinfo {author} {\bibfnamefont {R.~J.}\ \bibnamefont {Haug}},\ }\bibfield  {title} {\bibinfo {title} {Magnetoresistance induced by rare strong scatterers in a high-mobility two-dimensional electron gas},\ }\href {https://doi.org/10.1103/PhysRevB.90.165434} {\bibfield  {journal} {\bibinfo  {journal} {Phys. Rev. B}\ }\textbf {\bibinfo {volume} {90}},\ \bibinfo {pages} {165434} (\bibinfo {year} {2014})}\BibitemShut {NoStop}%
\bibitem [{\citenamefont {Wang}\ \emph {et~al.}(2016)\citenamefont {Wang}, \citenamefont {Samaraweera}, \citenamefont {Reichl}, \citenamefont {Wegscheider},\ and\ \citenamefont {Mani}}]{Wang2016}%
  \BibitemOpen
  \bibfield  {author} {\bibinfo {author} {\bibfnamefont {Z.}~\bibnamefont {Wang}}, \bibinfo {author} {\bibfnamefont {R.~L.}\ \bibnamefont {Samaraweera}}, \bibinfo {author} {\bibfnamefont {C.}~\bibnamefont {Reichl}}, \bibinfo {author} {\bibfnamefont {W.}~\bibnamefont {Wegscheider}},\ and\ \bibinfo {author} {\bibfnamefont {R.~G.}\ \bibnamefont {Mani}},\ }\bibfield  {title} {\bibinfo {title} {{Tunable electron heating induced giant magnetoresistance in the high mobility GaAs/AlGaAs 2D electron system}},\ }\href {https://doi.org/10.1038/srep38516} {\bibfield  {journal} {\bibinfo  {journal} {Scientific Reports}\ }\textbf {\bibinfo {volume} {6}},\ \bibinfo {pages} {38516} (\bibinfo {year} {2016})}\BibitemShut {NoStop}%
\bibitem [{\citenamefont {Samaraweera}\ \emph {et~al.}(2018)\citenamefont {Samaraweera}, \citenamefont {Liu}, \citenamefont {Gunawardana}, \citenamefont {Kriisa}, \citenamefont {Reichl}, \citenamefont {Wegscheider},\ and\ \citenamefont {Mani}}]{Samaraweera2018}%
  \BibitemOpen
  \bibfield  {author} {\bibinfo {author} {\bibfnamefont {R.~L.}\ \bibnamefont {Samaraweera}}, \bibinfo {author} {\bibfnamefont {H.-C.}\ \bibnamefont {Liu}}, \bibinfo {author} {\bibfnamefont {B.}~\bibnamefont {Gunawardana}}, \bibinfo {author} {\bibfnamefont {A.}~\bibnamefont {Kriisa}}, \bibinfo {author} {\bibfnamefont {C.}~\bibnamefont {Reichl}}, \bibinfo {author} {\bibfnamefont {W.}~\bibnamefont {Wegscheider}},\ and\ \bibinfo {author} {\bibfnamefont {R.~G.}\ \bibnamefont {Mani}},\ }\bibfield  {title} {\bibinfo {title} {{Coherent backscattering in quasi-ballistic ultra-high mobility GaAs/AlGaAs 2DES}},\ }\href {https://doi.org/10.1038/s41598-018-28359-0} {\bibfield  {journal} {\bibinfo  {journal} {Scientific Reports}\ }\textbf {\bibinfo {volume} {8}},\ \bibinfo {pages} {10061} (\bibinfo {year} {2018})}\BibitemShut {NoStop}%
\bibitem [{\citenamefont {Schluck}\ \emph {et~al.}(2018)\citenamefont {Schluck}, \citenamefont {Hund}, \citenamefont {Heckenthaler}, \citenamefont {Heinzel}, \citenamefont {Siboni}, \citenamefont {Horbach}, \citenamefont {Pierz}, \citenamefont {Schumacher}, \citenamefont {Kazazis}, \citenamefont {Gennser},\ and\ \citenamefont {Mailly}}]{Schluck2018a}%
  \BibitemOpen
  \bibfield  {author} {\bibinfo {author} {\bibfnamefont {J.}~\bibnamefont {Schluck}}, \bibinfo {author} {\bibfnamefont {M.}~\bibnamefont {Hund}}, \bibinfo {author} {\bibfnamefont {T.}~\bibnamefont {Heckenthaler}}, \bibinfo {author} {\bibfnamefont {T.}~\bibnamefont {Heinzel}}, \bibinfo {author} {\bibfnamefont {N.~H.}\ \bibnamefont {Siboni}}, \bibinfo {author} {\bibfnamefont {J.}~\bibnamefont {Horbach}}, \bibinfo {author} {\bibfnamefont {K.}~\bibnamefont {Pierz}}, \bibinfo {author} {\bibfnamefont {H.~W.}\ \bibnamefont {Schumacher}}, \bibinfo {author} {\bibfnamefont {D.}~\bibnamefont {Kazazis}}, \bibinfo {author} {\bibfnamefont {U.}~\bibnamefont {Gennser}},\ and\ \bibinfo {author} {\bibfnamefont {D.}~\bibnamefont {Mailly}},\ }\bibfield  {title} {\bibinfo {title} {{Linear negative magnetoresistance in two-dimensional Lorentz gases}},\ }\href {https://doi.org/10.1103/PhysRevB.97.115301} {\bibfield  {journal} {\bibinfo  {journal} {Phys. Rev. B}\ }\textbf {\bibinfo {volume} {97}},\ \bibinfo {pages} {115301}
  (\bibinfo {year} {2018})}\BibitemShut {NoStop}%
\bibitem [{\citenamefont {Horn-Cosfeld}\ \emph {et~al.}(2021)\citenamefont {Horn-Cosfeld}, \citenamefont {Schluck}, \citenamefont {Lammert}, \citenamefont {Cerchez}, \citenamefont {Heinzel}, \citenamefont {Pierz}, \citenamefont {Schumacher},\ and\ \citenamefont {Mailly}}]{Horn2021}%
  \BibitemOpen
  \bibfield  {author} {\bibinfo {author} {\bibfnamefont {B.}~\bibnamefont {Horn-Cosfeld}}, \bibinfo {author} {\bibfnamefont {J.}~\bibnamefont {Schluck}}, \bibinfo {author} {\bibfnamefont {J.}~\bibnamefont {Lammert}}, \bibinfo {author} {\bibfnamefont {M.}~\bibnamefont {Cerchez}}, \bibinfo {author} {\bibfnamefont {T.}~\bibnamefont {Heinzel}}, \bibinfo {author} {\bibfnamefont {K.}~\bibnamefont {Pierz}}, \bibinfo {author} {\bibfnamefont {H.~W.}\ \bibnamefont {Schumacher}},\ and\ \bibinfo {author} {\bibfnamefont {D.}~\bibnamefont {Mailly}},\ }\bibfield  {title} {\bibinfo {title} {Relevance of weak and strong classical scattering for the giant negative magnetoresistance in two-dimensional electron gases},\ }\href {https://doi.org/10.1103/PhysRevB.104.045306} {\bibfield  {journal} {\bibinfo  {journal} {Phys. Rev. B}\ }\textbf {\bibinfo {volume} {104}},\ \bibinfo {pages} {045306} (\bibinfo {year} {2021})}\BibitemShut {NoStop}%
\bibitem [{\citenamefont {Wang}\ \emph {et~al.}(2022)\citenamefont {Wang}, \citenamefont {Jia}, \citenamefont {Du}, \citenamefont {Pfeiffer}, \citenamefont {Baldwin},\ and\ \citenamefont {West}}]{Wang2022}%
  \BibitemOpen
  \bibfield  {author} {\bibinfo {author} {\bibfnamefont {X.}~\bibnamefont {Wang}}, \bibinfo {author} {\bibfnamefont {P.}~\bibnamefont {Jia}}, \bibinfo {author} {\bibfnamefont {R.-R.}\ \bibnamefont {Du}}, \bibinfo {author} {\bibfnamefont {L.~N.}\ \bibnamefont {Pfeiffer}}, \bibinfo {author} {\bibfnamefont {K.~W.}\ \bibnamefont {Baldwin}},\ and\ \bibinfo {author} {\bibfnamefont {K.~W.}\ \bibnamefont {West}},\ }\bibfield  {title} {\bibinfo {title} {{Hydrodynamic charge transport in an GaAs/AlGaAs ultrahigh-mobility two-dimensional electron gas}},\ }\href {https://doi.org/10.1103/PhysRevB.106.L241302} {\bibfield  {journal} {\bibinfo  {journal} {Phys. Rev. B}\ }\textbf {\bibinfo {volume} {106}},\ \bibinfo {pages} {L241302} (\bibinfo {year} {2022})}\BibitemShut {NoStop}%
\bibitem [{\citenamefont {Bockhorn}\ \emph {et~al.}(2024)\citenamefont {Bockhorn}, \citenamefont {Schuh}, \citenamefont {Reichl}, \citenamefont {Wegscheider},\ and\ \citenamefont {Haug}}]{Bockhorn2024}%
  \BibitemOpen
  \bibfield  {author} {\bibinfo {author} {\bibfnamefont {L.}~\bibnamefont {Bockhorn}}, \bibinfo {author} {\bibfnamefont {D.}~\bibnamefont {Schuh}}, \bibinfo {author} {\bibfnamefont {C.}~\bibnamefont {Reichl}}, \bibinfo {author} {\bibfnamefont {W.}~\bibnamefont {Wegscheider}},\ and\ \bibinfo {author} {\bibfnamefont {R.~J.}\ \bibnamefont {Haug}},\ }\bibfield  {title} {\bibinfo {title} {Influence of the electron density on the giant negative magnetoresistance in two-dimensional electron gases},\ }\href {https://doi.org/10.1103/PhysRevB.109.205416} {\bibfield  {journal} {\bibinfo  {journal} {Phys. Rev. B}\ }\textbf {\bibinfo {volume} {109}},\ \bibinfo {pages} {205416} (\bibinfo {year} {2024})}\BibitemShut {NoStop}%
\bibitem [{\citenamefont {Bartels}\ \emph {et~al.}(2026{\natexlab{b}})\citenamefont {Bartels}, \citenamefont {Horn-Cosfeld}, \citenamefont {Schluck}, \citenamefont {Cerchez}, \citenamefont {Mailly}, \citenamefont {Pierz}, \citenamefont {Schumacher}, \citenamefont {Sanvee}, \citenamefont {Horbach},\ and\ \citenamefont {Heinzel}}]{Bartels2026}%
  \BibitemOpen
  \bibfield  {author} {\bibinfo {author} {\bibfnamefont {F.}~\bibnamefont {Bartels}}, \bibinfo {author} {\bibfnamefont {B.}~\bibnamefont {Horn-Cosfeld}}, \bibinfo {author} {\bibfnamefont {J.}~\bibnamefont {Schluck}}, \bibinfo {author} {\bibfnamefont {M.}~\bibnamefont {Cerchez}}, \bibinfo {author} {\bibfnamefont {D.}~\bibnamefont {Mailly}}, \bibinfo {author} {\bibfnamefont {K.}~\bibnamefont {Pierz}}, \bibinfo {author} {\bibfnamefont {H.~W.}\ \bibnamefont {Schumacher}}, \bibinfo {author} {\bibfnamefont {B.}~\bibnamefont {Sanvee}}, \bibinfo {author} {\bibfnamefont {J.}~\bibnamefont {Horbach}},\ and\ \bibinfo {author} {\bibfnamefont {T.}~\bibnamefont {Heinzel}},\ }\bibfield  {title} {\bibinfo {title} {{Nonlinearities in the Hall resistance by retroreflective scattering}},\ }\href {https://doi.org/doi.org/10.1103/lg7z-nrjy} {\bibfield  {journal} {\bibinfo  {journal} {Phys. Rev. B}\ }\textbf {\bibinfo {volume} {113}},\ \bibinfo {pages} {165306} (\bibinfo {year} {2026}{\natexlab{b}})}\BibitemShut {NoStop}%
\end{thebibliography}
\end{document}